\pdfoutput=1
\documentclass{aa}  

\usepackage{txfonts}
\usepackage{ifpdf}              
\usepackage{url}                

\def\firstname{J.~M.}
\def\familyname{Corral-Santana}
\def\FileAuthor{\firstname~\familyname}
\def\FileTitle{\firstname~\familyname et al.}
\def\FileSubject{BlackCAT}
\def\FileKeyWords{X-ray binaries, Black Holes, BlackCAT, catalogue}

\usepackage{verbatim}            
\usepackage{textcomp}			 

\usepackage{latexsym}   
\usepackage{amssymb}    
\usepackage{amsfonts}   
\usepackage{amsmath}    
\usepackage{amstext}    

\usepackage{floatfig}

\usepackage{longtable}               
\usepackage{lscape}                  
\usepackage{multirow}                
\usepackage{multicol}                

\usepackage[table]{xcolor}           

\usepackage{footnote}                
\makesavenoteenv{tabular}
\usepackage{array}

\def\msun{\,$\rm M_\odot$}               
\def\m1{$M_1$}                           
\def\m2{$M_2$}                           
\def\k1{$K_1$}                           
\def\k2{$K_2$}                           

\def\ks{K{\rm s}}                        
\def\jtrece{\textit{Swift}\,J1357.2--0933}  
\def\j13{\textit{Swift}\,J1357.2--0933}  
\def\j18{XTE\,J1859+226}                 
\def\j04{XTE\,J1118+480}                 
\def\jcero{GRO\,J0422+32}                

\def\j11{XTE\,J1118+480}                 
\def\jonce{XTE\,J1118+480}               
\def\acero{A0620--003}                   
\def\a0{A0620--003}                      
\def\v4{V404\,Cyg}                       

\def\maxi{MAXI\,J1659--152}              

\def\j16{XTE\,J1650--500}                
\def\j15{XTE\,J1550--564}                

\def\deg{\mbox{$^{\circ}$}}              
\def\kms{\,$\rm km\,s^{-1}$}             
\def\arcsec{\hbox{$^{\prime\prime}$}}    
\def\arcmin{\hbox{$^{\prime}$}}          

\def\xe #1#2{$#1\times10^{#2}$}
\def\xebf #1#2{$\mathbf{{#1}\times10^{#2}}$}

\newcommand{\Tab}{{Table~}}     
\newcommand{\equ}{{eq.~}}       
\newcommand{\fig}{{Fig.~}}      
\newcommand{\tab}{{Table~}}     

\newcommand{\sect}{{Sec.~}}    

\newcommand\fnm{\,\footnotemark} 

\newcommand{\txw}{\textwidth}
\newcommand{\txh}{\textheight}

\def\nar{\aaref@jnl{New Astron. Rev.}}       
\def\actaa{\aaref@jnl{Acta. Astron.}}        

\usepackage{rotating}

\usepackage{natbib,twoopt}
\bibpunct{(}{)}{;}{a}{}{,} 
\makeatletter
\newcommandtwoopt{\citeads}[3][][]{\href{http://adsabs.harvard.edu/abs/#3}%
{\def\hyper@linkstart##1##2{}%
\let\hyper@linkend\@empty\citealp[#1][#2]{#3}}}
\newcommandtwoopt{\citepads}[3][][]{\href{http://adsabs.harvard.edu/abs/#3}%
{\def\hyper@linkstart##1##2{}%
\let\hyper@linkend\@empty\citep[#1][#2]{#3}}}
\newcommandtwoopt{\citetads}[3][][]{\href{http://adsabs.harvard.edu/abs/#3}%
{\def\hyper@linkstart##1##2{}%
\let\hyper@linkend\@empty\citet[#1][#2]{#3}}}
\newcommandtwoopt{\citeyearads}[3][][]%
{\href{http://adsabs.harvard.edu/abs/#3}
{\def\hyper@linkstart##1##2{}%
\let\hyper@linkend\@empty\citeyear[#1][#2]{#3}}}
\makeatother

\usepackage{breakurl}

\ifpdf                          
  \usepackage[baseurl=http://, 
   pdfpagemode=UseNone, pdfstartview=XYZ, pdfstartpage=1,
   colorlinks=true, 
   linkcolor=blue,   
   citecolor=blue,   
   urlcolor=red,    
   breaklinks=true,
   ]
   {hyperref}        
  \hypersetup{
    pdfauthor   = \FileAuthor,    
    pdftitle    = \FileTitle,     
    pdfsubject  = \FileSubject,   
    pdfkeywords = \FileKeyWords}  

\else                            
  \usepackage[dvips,breaklinks=true]{hyperref}
\fi

\begin{document}

   \title{BlackCAT: A catalogue of stellar-mass black holes in X-ray transients}
   
   \author{J.~M. Corral-Santana
          \inst{1}\fnmsep\thanks{jcorral@astro.puc.cl}
          \and
          J. Casares\inst{2,3}
          \and
          T. Mu\~noz-Darias\inst{2,3}
          \and
          F.~E. Bauer\inst{1,4,5}
          \and
          I.~G. Mart\'inez-Pais\inst{2,3}
          \and
          D.~M. Russell\inst{6}
          }

   \institute{Instituto de Astrof\'{\i}sica, Facultad de F\'{i}sica, Pontificia Universidad Cat\'{o}lica de Chile (IA-PUC), Casilla 306, Santiago 22, Chile
        \and
             Instituto de Astrof\'isica de Canarias (IAC), V\'ia L\'actea, s/n, E-38205 La Laguna, S/C de Tenerife, Spain
         \and
             Departamento de Astrof\'isica - Universidad de La Laguna (ULL), E-38206 La Laguna, S/C de Tenerife, Spain
         \and
             Millennium Institute of Astrophysics (MAS), Nuncio Monse\~{n}or S\'{o}tero Sanz 100, Providencia, Santiago de Chile
         \and
             Space Science Institute, 4750 Walnut Street, Suite 205, Boulder, Colorado 80301
         \and
             New York University Abu Dhabi, PO Box 129188, Abu Dhabi, UAE\\
             }

   \date{}

 
  \abstract
   {}
   {During the last $\sim50$ years, the population of black hole candidates in X-ray binaries has increased considerably with 59 Galactic objects detected in transient low-mass X-ray binaries, plus a few in persistent systems (including $\sim5$ extragalactic binaries). }
   {We collect near-infrared, optical and X-ray information spread over hundreds of references in order to study the population of black holes in X-ray transients as a whole.
   }
   {We present the most updated catalogue of black hole transients, which contains X-ray, optical and near-infrared observations together with their astrometric and dynamical properties. It provides new useful information in both statistical and observational parameters providing a thorough and complete overview of the black hole population in the Milky Way. Analysing the distances and spatial distribution of the observed systems, we estimate a total population of $\sim1300$ Galactic black hole transients. This means that we have already discovered less than $\sim5$\% of the total Galactic distribution. The complete version of this catalogue will be continuously updated online and in the Virtual Observatory, including finding charts and data in other wavelengths.
}
   {}

   \titlerunning{BlackCAT}
   \authorrunning{Corral-Santana et al.}

   \maketitle
%

\section{Introduction}
\label{sec:intro}
  X-ray binaries (XRBs) are systems formed by either a neutron star (NS) or a black hole (BH), which is accreting mass from a companion donor star. Their detection began thanks to the development of space-based instrumentation in the 1960's, with the number of detections rising substantially since the implementation of all-sky monitors on-board X-ray satellites, e.g. Ginga (1987--1991), RXTE (1996--2012) and, more recently, \textit{Swift} (2004) and MAXI (2009). They are broadly divided into high-mass X-ray binaries (HMXBs) and low-mass X-ray binaries (LMXBs) according to the mass of the donor star. In the former, the early spectral type (O--B) massive star ($\geq10$\msun) transfers material to the compact object mainly through strong stellar winds. On the other hand, in LMXBs the K--M spectral type star ($M_2\leq1$\msun) fills the Roche lobe and transfers mass by Roche lobe overflow through the inner Lagrangian point \citep{Charles2006}.
  In this latter case, the material forms an accretion disc around the compact object whereby material is accreted. A few XRBs are also found with intermediate mass companions of spectral types in the range B--F. These so-called intermediate-mass XRBs (IMXBs) have been proposed to be the progenitors of some LMXBs through an episode of enhanced mass-transfer rate \citep{Podsiadlowski2002}.
  
  The mass transfer rate ($\dot{M}$) largely determines the observational properties and gives rise to a sub-classification within the LMXBs class. \textit{Persistent} sources are those with high accretion rates \citep[$\dot{M} \sim 10^{-10}$\msun~yr$^{-1}$;][]{Tanaka1996} and X-ray luminosities close to the Eddington limit. This high luminosity ensures that the outer parts of the accretion disc dominate the optical spectrum through reprocessing and effectively hide the companion star. On the contrary, \textit{transient} sources are systems with low accretion rates \citep[$\dot{M} \leq 10^{-9}$\msun~yr$^{-1}$;][]{Tanaka1996} which exhibit long quiescent states and sporadic outburst episodes produced by thermal-viscous instabilities in the accretion disc (see \citealt{Frank2002} for further explanation). During outbursts, the brightness of the system rises to similar luminosities as those found in the \textit{persistent} sources. After the outburst, \textit{transients} decay back to quiescent states where they remain for most of their lifetimes. Typical recurrence times between outburst span from years to centuries depending on $\dot{M}$ \citep{Ritter2002,McClintock2006a}.

  Observations have revealed that $\sim25$\% of the \textit{transient} LMXBs contain bursting NS \citep{King1996} while the rest ($\sim75$\%) display X-ray spectral and/or timing properties characteristic of accreting black holes (hereafter we will refer to the transient black hole systems as black hole transients or BHTs). In this paper we focus on the properties of the Galactic BHTs as they represent the vast majority of the population of BHs. However, there are a few persistent, non-active and extragalactic XRBs that harbour or may contain BHs.
  
    Regarding the extragalactic population of BHs, dynamical evidence has been presented for LMC~X-1 \citep[a $11\pm1$\msun~BH with an O7III companion;][]{Orosz2009}, LMC~X-3 (a $7.0\pm0.6$\msun~BH with a B3--5V star; \citealt{Orosz2014} and \citealt{Val-Baker2007}) and M33~X-7 \citep[a $16\pm1$\msun~BH with a O7--8III star;][]{Orosz2007}, the first eclipsing stellar-mass BH ever detected.\\
    On the other hand, indirect evidence for the presence of BHs in the HMXBs NGC\,300~X-1 \citep[$12-24$\msun~BH with a Wolf-Rayet star;][]{Crowther2010} and IC\,10~X-1 \citep[a $23-34$\msun~BH with a Wolf-Rayet star;][]{Silverman2008} has been postulated. Owing to the unique challenges of observing the wind-dominated Wolf-Rayet companions, the masses of the compact objects in these systems are rather uncertain and indeed a neutron star cannot be ruled out \citep{Binder2015,Laycock2015}.
    Finally, we should also mention ultraluminous X-ray sources (ULXs), systems with X-ray luminosities greater than the Eddington limit for a $10$\msun~BH. The source of these luminosities is still uncertain and it has been proposed that they may be produced by intermediate-mass  BHs \citep[$\sim10^3$\msun. See e.g. ][]{Miller2003} or stellar-mass BHs \citep[e.g., ][]{Poutanen2007,Kawashima2012}. More recently, it has been found for one case that the compact object is a NS \citep[M82 X-2;][]{Bachetti2014} confounding our understanding of ULXs even further.  
  
  Hereafter, we will only focus on the Galactic population of BHs. We have performed a thorough search of all X-ray systems published in literature since 1962, when the first extrasolar X-ray source was detected, \citep[the NS system Sco X-1; ][]{Giacconi1962} up to 2015. In \sect\ref{sec:cat} we motivate this catalogue with a historical view of the sample of BHTs and present the catalogue itself. In \sect\ref{sec:anal} we study the population of BHTs, where we analyse the vertical distribution of BHTs constrain the expected number of systems in the Milky Way. 
  In \sect\ref{sec:dynamical} we focus on the population of dynamically confirmed BHs, presenting their distributions of periods, magnitudes and masses.

   \begin{figure}
   \centering
   \includegraphics[width=0.45\txw,height=0.2\txh]{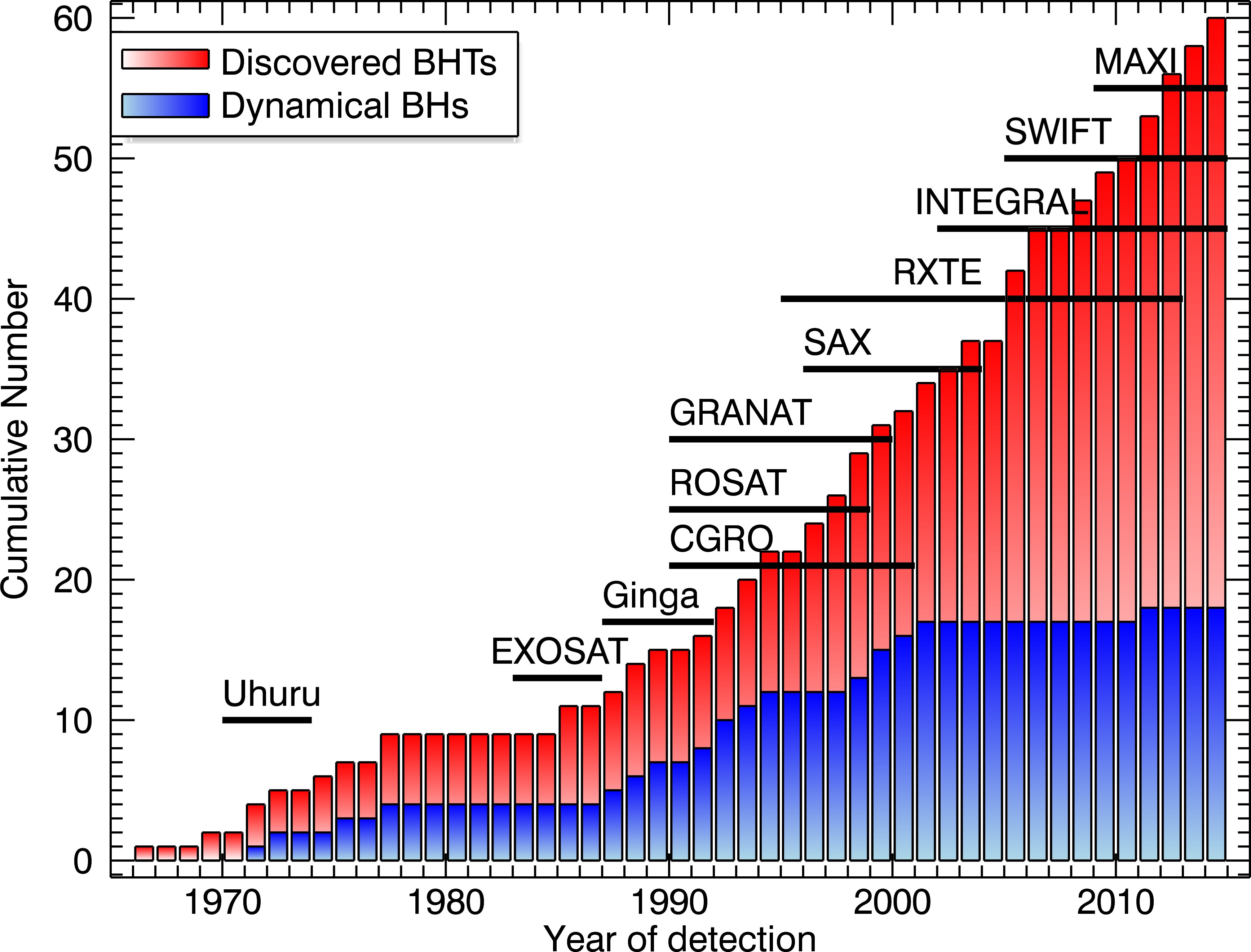}
   \caption{Cumulative histogram of discovered (red) and dynamically confirmed (blue) 
            BHTs as a function of time. Here, we also count \jtrece~as a dynamical BH. The lifetimes of 
            the main X-ray satellites with all-sky monitor capabilities are shown in black lines.}
              \label{fig:histobhs}
    \end{figure}

\begin{center}
  \begin{figure}[h!!!!!!]
   \centering
    \includegraphics[width=0.45\txw]{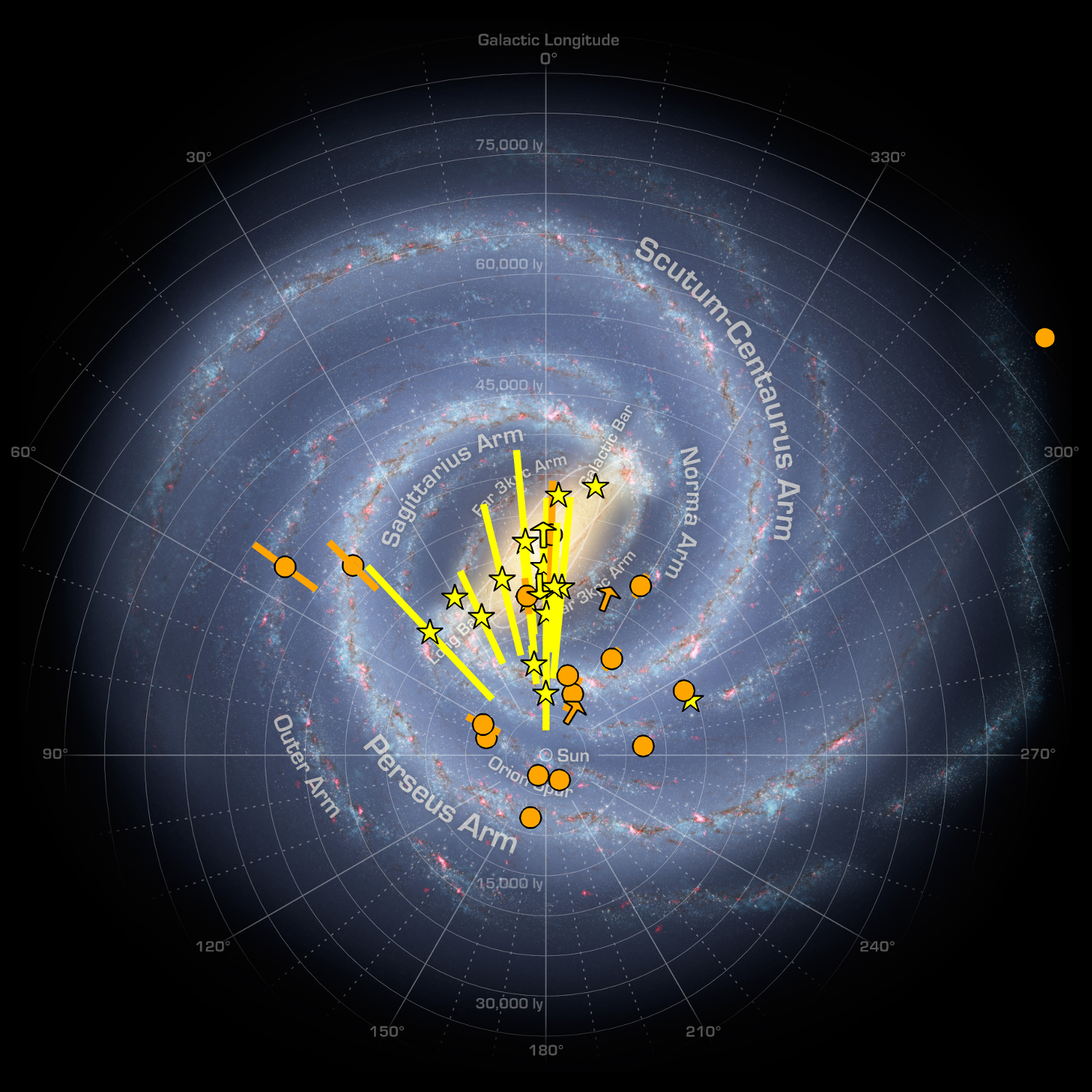}
    \caption[Distribution of BHs in the Galaxy. Polar view]{The Galactic distribution of 35 BHTs with distance estimates (\Tab\ref{tab:census1}) as seen from the Galactic pole. Dynamically confirmed black holes are marked in orange circles) whereas BH candidates are indicated by yellow stars. Distance ranges are represented with bars while the systems with lower limits are marked with arrows, following the same colour index. Some systems are overlapped with other. Each faint white ring in the background image represents an additional $\sim1.5$\,kpc from the Sun. Objects seem to concentrate along the direction towards the Galactic centre. The dynamical BHT in the upper right of the figure is BW~Cir. Background image credit: NASA/JPL-Caltech/R. Hurt (SSC/Caltech). 
    }
    \label{fig:bhs_mw_out}
  \end{figure}
\end{center}

\section{BlackCAT: The catalogue of Galactic BHs}
\label{sec:cat}
  Since 1966 up to 2015, 59 BHTs have been discovered, represented in \fig~\ref{fig:histobhs} as a cumulative histogram of red bars, the slope of which represents a detection rate of $\sim1.2$ targets per year. However, only 17 of these have been dynamically confirmed to harbour accreting BHs (i.e. mass function $\gtrsim3$\msun; see, e.g., \citealt{Casares2014a}) and are represented by the blue bars in \fig~\ref{fig:histobhs}. Here, we should also add \jtrece, where the dynamical confirmation is indirect because it is not based on the detection of the secondary star. However, it has robust evidence of containing a BH (see \citealt{Corral-Santana2013} and \citealt{MataSanchez2015} for more details).
  
  Therefore, the 17+1 dynamical BHs represent $\sim30$\% of the total number of BHTs discovered so far. This low fraction is due to the fact that most BHTs become too faint in quiescence for radial velocity studies with current instrumentation due to intrinsically faint companions, high extinction, large distance or a combination of these. This problem could be alleviated with the exquisite sensitivity of future facilities like the E-ELT, specially with better IR instrumentation. The rest of the population of BHTs are named \textit{BH candidates} because they share similar X-ray characteristics in outburst to the confirmed ones, but they lack a final dynamical confirmation (see, e.g., \citealt{McClintock2006} and \citealt{Belloni2011} for reviews on the X-ray observational properties of the BH candidates). Indeed, NS and BH sometimes display qualitatively similar phenomenology in outburst \citep[see, e.g.,][]{Munoz-Darias2014} and we cannot discard that some of the objects included in this catalogue might harbour NS.
  
  BlackCAT is a complete catalogue containing the astrometric, photometric (near-infrared -NIR-/optical magnitudes in outburst and quiescence), number of outbursts, the peak X-ray flux, distance, finding charts and dynamical parameters of all the BHTs discovered so far. The electronic and most complete version of this catalogue is available on \url{www.astro.puc.cl/BlackCAT} although all data will also be available through the Virtual Observatory. In this paper we present the most relevant properties divided in the tables explained below.   
  
 We divide the catalogue in 3 main types of Galactic BHs: transients, persistent and non-active, depending on their X-ray activity.

  \subsubsection*{Persistent}
    Cyg~X-1 \citep[a $15\pm1$\msun~BH with a O9.7\,Iab donor star;][]{Orosz2011} is the only confirmed Galactic BH in a persistent X-ray binary. On the other hand, 4U\,1957+11 \citep{Wijnands2002b,Nowak2008,Nowak2012,Hakala2014} and 1E\,1740.7-2942 \citep{Sunyaev1991b} are persistent BH candidates but they have not been dynamically confirmed. GRS~1758-258 \citep{Mandrou1990} is a quasi-persistent microquasar with a large extinction A$_v\sim 8.4$ \citep{Mereghetti1997} located near the Galactic center region. Finally, SS~443 \citep{Stephenson1977} is a non-transient source with a supercritical accretion regime onto a relativistic star (see \citealt{Fabrika2004} for a detailed review on the system). It is very likely a \textit{BH candidate} with indirect arguments that support a compact object of $10-20$\msun~with a high inclination \citep{Eikenberry2001a}. However, despite it has been intensively studied for almost 30 years, the nature of this system is still uncertain and it could be a Galactic ULX.

  \subsubsection*{Non-active BHs}    
     Further, MWC\,656 has been recently proposed as the first BH HMXB with a Be-type companion star \citep[$3.8-6.9$\msun~BH with a B1.5-2\,III star;][]{Casares2014}. This is based on radial velocity curves of gas encircling the companion star and the spectroscopic mass of the Be star. This system has not shown any type of outbursting activity, so we label it as a non-active BH. There are other 3 black hole candidates discovered in globular clusters: the flat-spectrum radio sources M22-VLA1 and M22-VLA2 \citep{Strader2012} and M62-VLA1 \citep{Chomiuk2013}.\\
  
   \subsubsection*{Transients}
   The rest of systems are transients, although some of them could be considered as semi-persistents (e.g. GRS\,1915+105 or 4U\,1630-472) because of their long stays in the outburst state.
    Below, we detail the content of the tables presented in this paper. The systems included are either dynamically confirmed BHs or have shown spectra and/or timing features typically found in BHs \citep{Belloni2011}. We also note the existence of Cyg~X-3 \citep{Giacconi1967}, a transient source showing strong radio outbursts. However, the nature of the primary accompanying the Wolf-Rayet donor is unclear and it could be either a NS or a BH. Thus, we have excluded Cyg~X-3 from our list of BHTs.\\
   
\textbf{\tab\ref{tab:census1}: Astrometric properties} presents the basic astrometric properties of the BHTs sorted chronologically by year of detection in X-rays. The dynamically confirmed BHs are highlighted in grey. The columns distribution is:\par
    (1) Year of discovery of the BHT;\par
    (2) Name of the system and optical counterpart when known;\par
    (3-5) Right ascension (RA) and declination (DEC) coordinates in equinox J2000. The accuracy in the astrometry and the source of the coordinates are also shown;\par
    (6-7) Galactic longitude ($\ell$) and latitude ($b$) in degrees;\par
    (8-9) Estimated distance ($d$) and height above the Galactic plane ($z$) in kpc;\par
    (10) Number of outbursts detected after discovery in X-rays;\par
    (11) References for the detection, best coordinates and distance determinations.\\ \par
    
\textbf{\tab\ref{tab:census2}: Peak X-ray flux and optical/NIR photometric parameters} shows the main properties both in outburst and in quiescence for all the BHTs presented in \tab\ref{tab:census1}. We also list the measured or predicted orbital period. Again, the dynamically confirmed BHs are highlighted in grey. The column distribution is:\par
   (1) ID number used for cross-reference with the web version of this catalogue;\par
   (2) Name of the system and optical counterpart when known;\par
   (3) Peak X-ray flux in erg~s$^{-1}$~cm$^2$, standardized to the 2--10\,keV band. To do so, we begin with the X-ray flux published in the literature (or in archive). We assume a power-law spectrum with a photon index $\Gamma=2$ \citep{Belloni2011} and the total neutral Galactic Hydrogen column density ($N_{\rm H}$) published by \cite{Kalberla2005}. If there is a measured $N_{\rm H}$ published in literature derived from direct X-ray spectral analysis, we use this instead of the radio-derived interstellar one; \par
   (4-5) Optical or IR magnitude in the peak of the outburst and quiescence, respectively, in the AB system. In order to document the original observed band, we provide the name of the band in its original system;\par 
   (6) Optical Galactic extinction [$E(B-V)$] reported in the literature. If unknown, we list the total Galactic line-of-sight absorption given by the \cite{Schlafly2011} dust maps.\par
   (7) Reported or estimated orbital period of the binary in hours;\par
   (8) References for all the parameters above.\\ \par

\textbf{\tab\ref{tab:census3}: The dynamical BHs: Photometric parameters in quiescence} lists the optical/NIR photometry of the dynamical BHTs in quiescence~\par
   (1) Preferred name of the dynamically confirmed BHT;\par
   (2-9) Quiescent magnitudes of the dynamically confirmed BHTs. All the magnitudes were transformed to the AB system using the transformation coefficients taken from table 2 in \cite{Frei1994} and equations 5 in \cite{Blanton2005}. In order to document the original observed band, we provide the name in its original system.\par
   (10) References of the magnitudes.\\ \par

\textbf{\tab\ref{tab:census4}: The dynamical BHs:  Binary parameters} provides the dynamical parameters of the BHTs \par
   (1) Preferred name of the dynamically confirmed BHT;\par
   (2) Spectral type of the companion star;\par
   (3) The orbital period of the binary in hours;\par
   (4) The radial velocity of the companion star ($K_2$) in \kms;\par
   (5) The mass function of the BH $f(M_1)$ in \msun;\par
   (6) The mass of the BH ($M_1$) in \msun. If there is an uncertain value, we preferred to show a range of masses;\par
   (7) The binary mass ratio $q=M_2/M_1$;\par
   (8) The inclination of the system ($i$) in degrees;\par
   (9) The rotational broadening ($v_{rot}\sin i$) in \kms;\par
   (10) References of all the parameters.\\ \par

\begin{figure*}[!htp]
 \centering
    \includegraphics[width=0.8\txw]{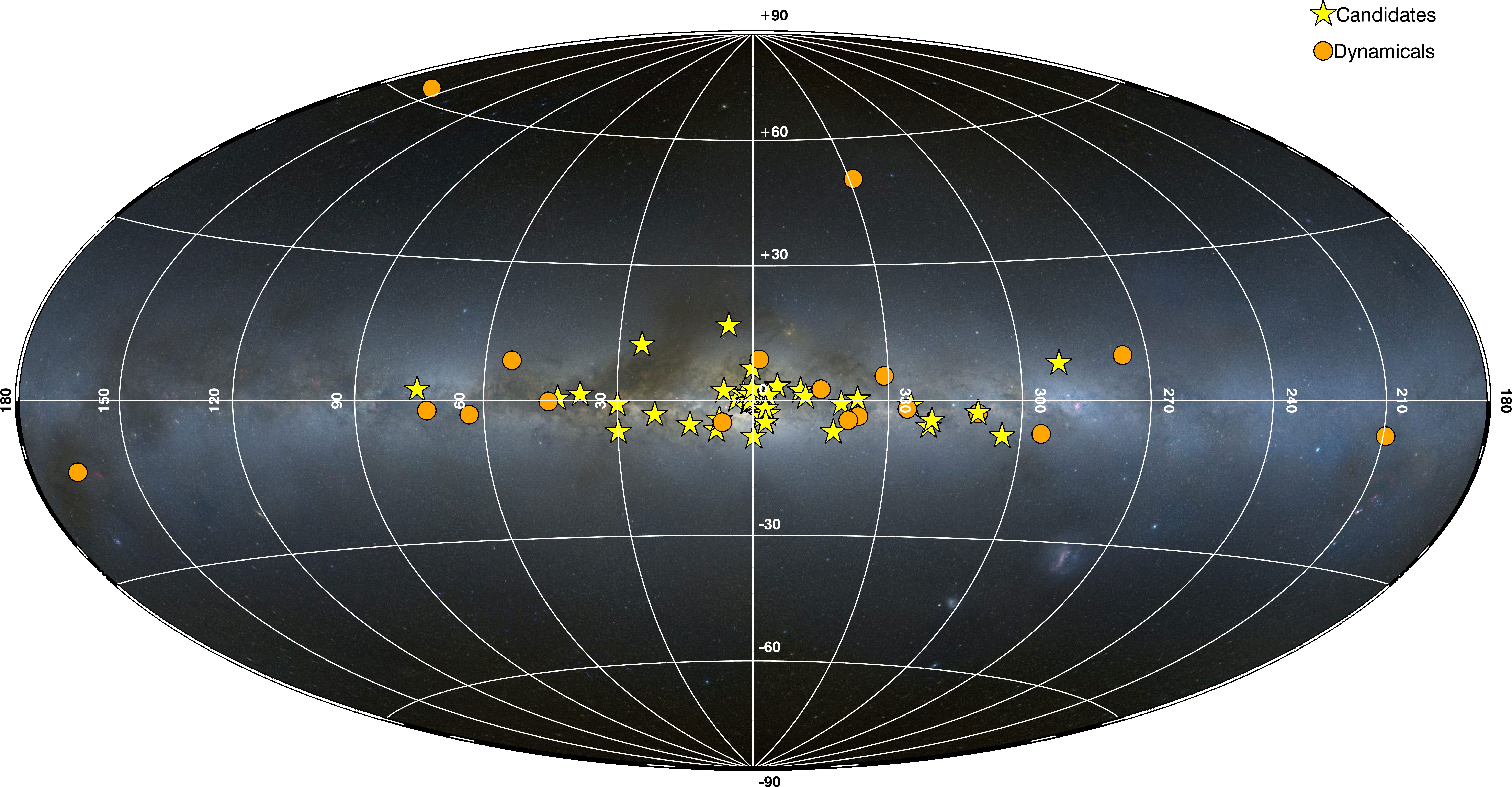}
    \includegraphics[width=0.49\txw,height=0.2\txh]{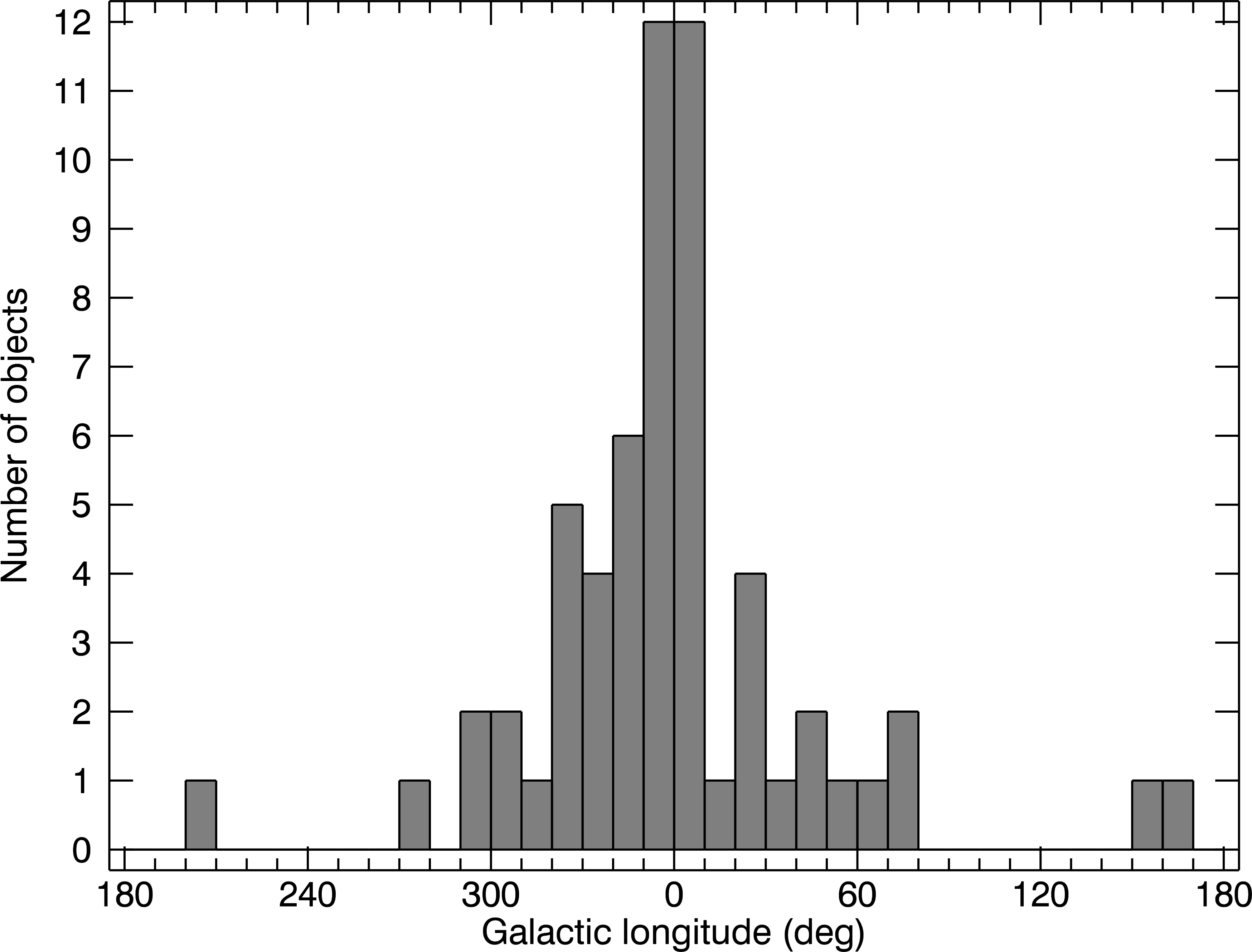}
    \includegraphics[width=0.49\txw,height=0.2\txh]{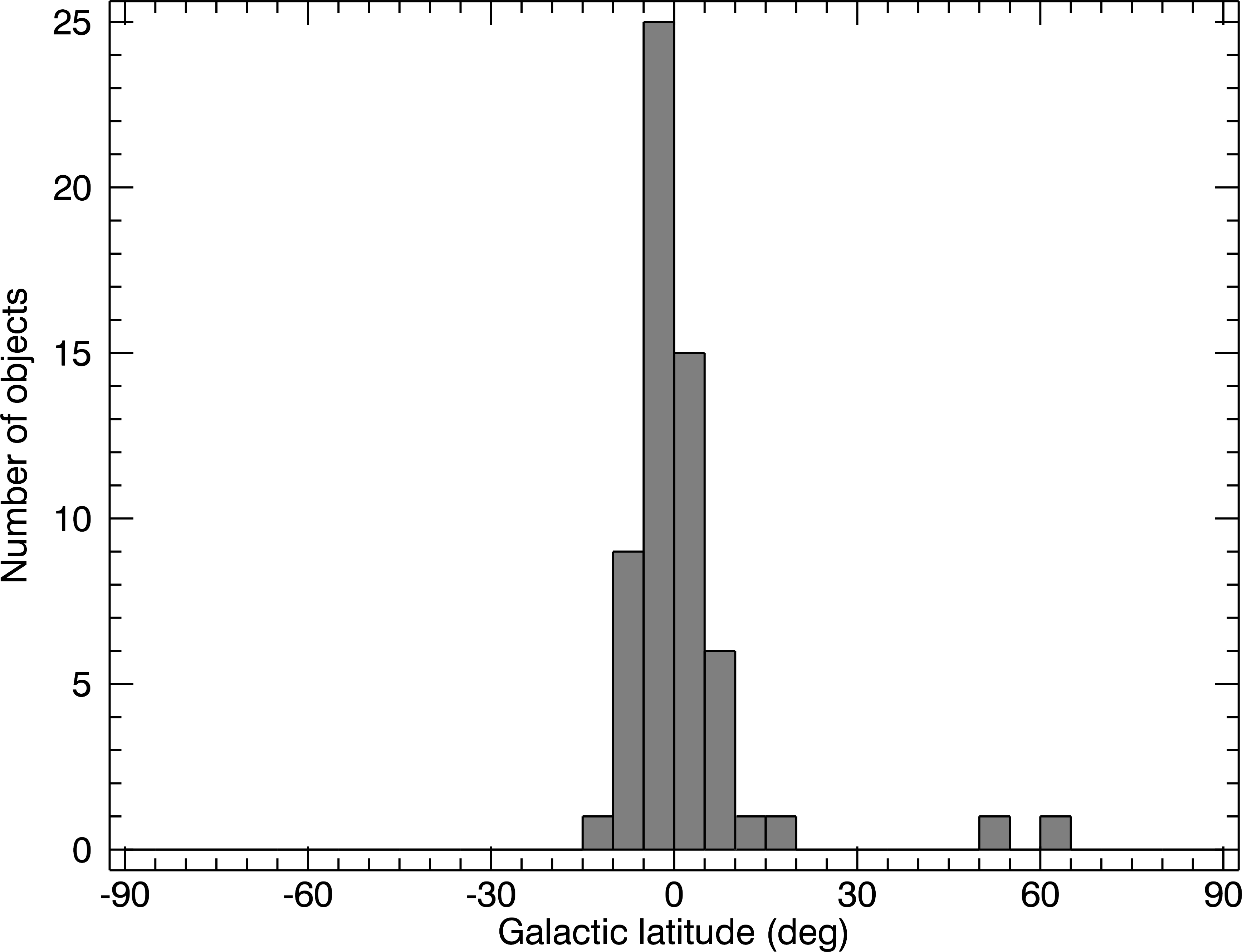}
    \caption[Spatial distribution of BHs]{\textit{Top}: The distribution of dynamically confirmed BHs (circles) and BH candidates (stars) in the Galactic plane, using the Hammer projection. \cite[Background image credit:][]{Mellinger2009}. Some of the symbols are overlapped by other, specially in the Galactic centre region. \textit{Bottom}: Histogram of the distribution of BH transients in Galactic longitude (\textit{left}) and latitude (\textit{right}). A 10\deg~bin size in longitude and 5\deg~bin in latitude were used respectively.     }
    \label{fig:bhs_mw_plane}
    \label{fig:histo_gal}
\end{figure*}
  
At the end of each table, we present a detailed list with the particularities marked in \tab\ref{tab:census1} and \tab\ref{tab:census2}. 
 
This census is the most updated available and, compared to previous catalogues (e.g. \citealt{Gottwald1991,Chen1997,Remillard2006,Ritter2003,Liu2006,Liu2007}), it represents a substantial improvement in both statistics and observational parameters. It is focused only in the population of BHs but providing a thorough and more complete coverage of optical and NIR data and dynamical parameters. It is online and will be continuously updated with more systems and information in other spectral bands.

\section{Analysis of the spatial distribution of BHTs}
\label{sec:anal}
  \tab\ref{tab:census1} allows us to perform a statistical study of the distribution of BHTs in the Galaxy. In \fig\ref{fig:bhs_mw_out}, we plot the 35 objects with estimated distances as viewed from the pole of the Milky Way. Filled orange circles represent the dynamically confirmed BHs while yellow stars mark the BH candidates. About 50\% of the confirmed BHs are located within 4.5\,kpc of the Sun. This is a clear indication that interstellar extinction is a severe limitation to dynamical mass determinations. Moreover, from \fig\ref{fig:bhs_mw_out} it seems that almost all BHs lie within a spiral arm. Thus, for BHTs with uncertain distances, likely values could be estimated or constrained using the distance to the spiral arm.
  
  \fig\ref{fig:bhs_mw_plane} shows the position of all the BHTs in Galactic coordinates overlaid on a projected image of the Milky Way, as well as histograms of their distributions in Galactic longitude and latitude. There is a clear concentration of objects towards the direction of the bulge ($340^\circ<\ell<20^\circ$ and $|b|<10$\deg, \citealt{Dwek1995}) and disc.  There are only two objects located at high Galactic latitudes (\jonce~at $b=+62$\deg~and \jtrece~at $b=+50$\deg), while three BHTs lie between $150^\circ<\ell<210^\circ$. 31 out of the 59 BHTs have no reported quiescent optical counterparts. In addition, eighteen of them were detected in outburst, but they are either too faint in quiescence or the crowding of the field has prevented the detection of a quiescent counterpart. On the other hand, 28 radio counterparts have been found (mostly during outburst). In column 5 of \tab\ref{tab:census1} we indicate the range of the source of the coordinates together with the accuracy.

  Stellar evolution predicts a population of $10^8 - 10^9$ BHs in our Galaxy \citep{vandenHeuvel2001,Remillard2006}. However, the total number of Galactic BHTs is very uncertain and depends on several considerations. \cite{vandenHeuvel1992,vandenHeuvel2001} estimated a population between of $500$ and $1000$ BHTs based on only 3-4 BHTs known at the time. Current estimates predict $10^3-10^4$ Galactic BHTs \citep{Romani1998,Kiel2006,Yungelson2006} based on models, although these numbers are likely underestimated because of the existence of systems with extremely long outburst duty cycles \citep{Ritter2002} or very faint peak X-ray luminosities \citep{King2006}.

  Here we want to revisit the problem of the size of the Galactic population of BHs, based on the currently observed sample, which is heavily affected by extinction. \cite{Duerbeck1984} modelled the density distribution of several interacting binaries in the Galaxy. In a similar way, we want to obtain the density of Galactic BHTs $\rho(z)$ and analyse the height distribution ($z$; column 8 in \tab\ref{tab:census1}). Only 35 out of the 59 detected BHTs have estimated distances. 
  Due to their transient nature, we have only detected those systems which went into outburst. Therefore, in a given period of time, we have discovered only a fraction of them (e.g., the brightest and/or closest BHTs and with the shortest outburst cycles). Thus, in order to derive the true $z$ distribution we first need to find the maximum radial distance out to which completeness of the sample is guaranteed, i.e., such that none of the X-ray outbursts were missed.

\begin{figure}[!htp]
 \centering
    \includegraphics[width=0.5\txw,height=0.3\txh]{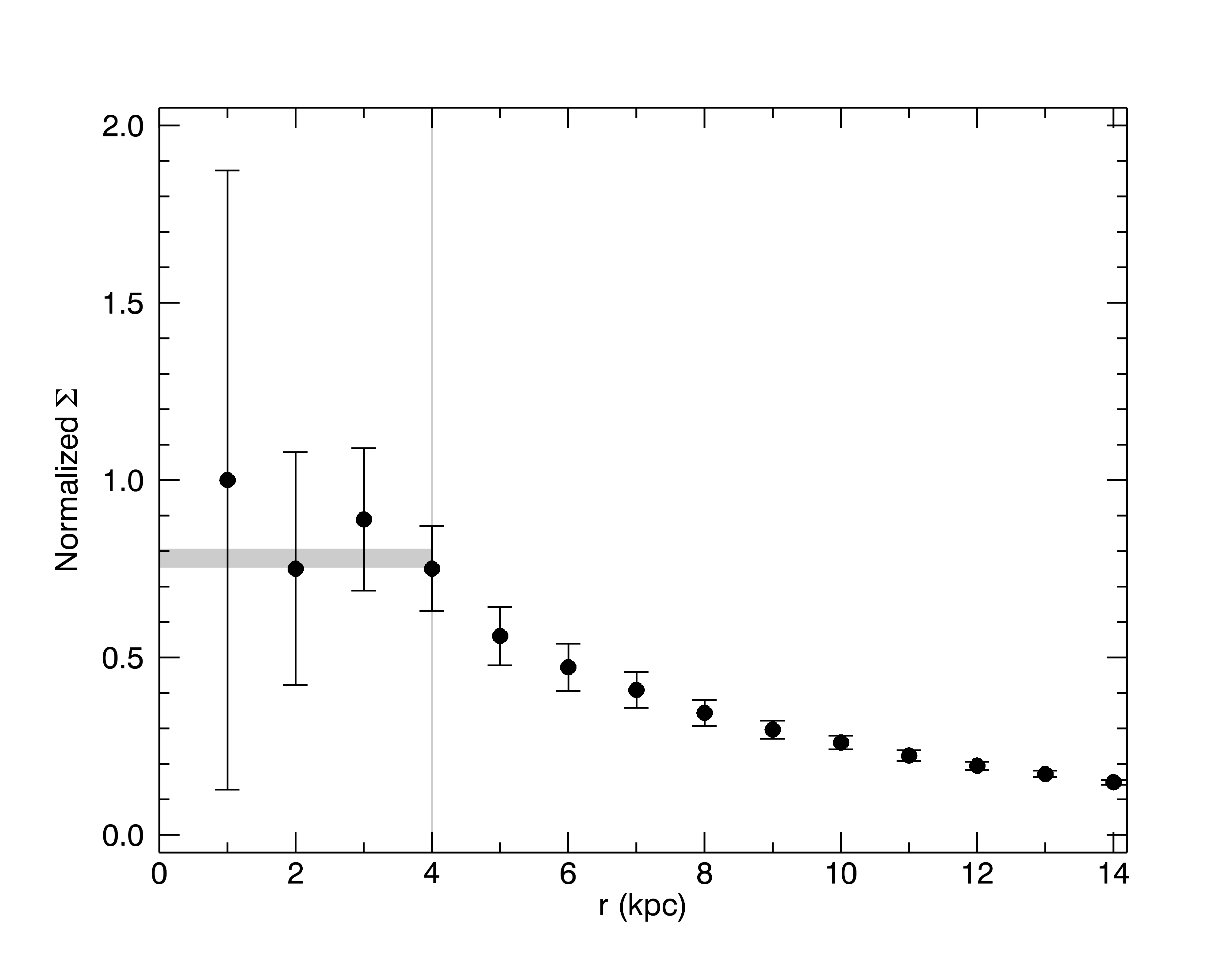}
    \caption[Completeness of the sample of BHTs]
            {Variation of the density of BHTs ($\Sigma$) in a cylindrical volume of infinite height $h$ with increasing radii $r$. Above $r>4$\,kpc the density decreases and therefore we infer the sample of BHTs is complete out to $r\sim4$\,kpc (grey bar).}
    \label{fig:complet}
\end{figure}

The completeness of the sample can be obtained from the analysis of the density of objects as a function of the radial distance projected on the Galactic plane ($r$). The latter is derived from the Galactic latitude and distance listed in columns 5 and 6 of \tab\ref{tab:census1}. 

Here, we have used 31 out of the 35 BHTs with a reported distance (we exclude the 4 systems with lower or upper limits). In addition, we consider a conservative 50\% uncertainty in systems with only a rough estimate of distance. To derive the density of objects ($\Sigma$) up to a radii $r$, we count the number of BHTs lying in cylinders or radius $r$ centred at the Sun and infinite height ($h$). However, given the large uncertainties with distance, in some cases an object could be placed in more than a single cylinder within errors. To account for this effect, for each object we randomly generate 10000 values assuming a Gaussian distribution. Here, we naively assume the distance errors follow this sort of distribution, under the assumption that all of the errors that went into the distance errors were Gaussian. In reality, we have no easy way to know this, as the values are from the literature, determined by many groups and in many cases not fully documented. Thus, the final density of objects obtained in a cylinder is given by the median value of all the 10000 $\Sigma$ obtained in the process, with the error given by the standard deviation. 
The result is shown in \fig\ref{fig:complet} where the density is represented as a function of the radial distance and normalized to its maximum value. In this figure we find a plateau in $\Sigma$ up to $r=4$\,kpc (within errors) and a decrease for $r>4$\,kpc. We interpret this as an indication that the density of objects is approximately constant with increasing radii up to $\sim4$\,kpc. 
We should note here that it is difficult to establish a limit, but the density decreases clearly at $r>4$\,kpc whereas it is not clear before that. Moreover, using the counting method proposed by \cite{Duerbeck1984}, we obtain a more abrupt decrease in $\Sigma$ for $r>4$\,kpc. However, we believe this technique is not suitable for systems with large uncertainties in the distance.\\
We also perform a complementary analysis based on the X-ray luminosity. The least luminous BHT detected is \jonce~which reached an X-ray luminosity of \xe{3.6}{35}\,$\rm erg\,s^{-1}$ in the 2-10\,keV range \citep{Dunn2010}. Therefore, we can assume that no BHT peaked at a luminosity lower than that. The sensitivity limits by the all-sky monitor (ASM) on RXTE in the 2-10\,keV range have been above \xe{2.4}{-10}$\,\rm erg\,s^{-1}\,cm^{-2}$ since 1996, which means that any BHT that reached \xe{3.6}{35}\,$\rm erg\,s^{-1}$ within 3.5\,kpc would have been detected (without considering absorption). This is consistent with the method described above and, therefore, we assume that the sample of BHTs is complete out to $\sim4$\,kpc.

\begin{table}
 \caption{List of BHTs within $r=4$\,kpc of completeness.}
 \label{tab:complete}
 \centering
 \scriptsize{
  \begin{tabular}{ll c c l}         
  \hline\hline                     
   ID & Name & $d$   & $z$   & Ref.\\    
       &  & (kpc) & (kpc) &    \\ 
   \hline                           
     40&XTE\,J1818-245 & $3.55\pm0.75$& $-0.26\pm0.05$ & \cite{CadolleBel2009} \\ 
     33 &XTE\,J1650-500 & $2.6\pm0.7$  & $-0.15\pm0.04$ & \cite{Homan2006} \\ 
     32 &XTE\,J1118+480 & $1.7\pm0.1$  & $1.52\pm0.09$  & \cite{Gelino2006} \\ 
     21 &GRO\,J1655-40  & $3.2\pm0.2$  & $0.13\pm0.01$  & \cite{Hjellming1995} \\ 
     20 &GRS\,1716-249  & $2.4\pm0.4$  & $0.29\pm0.05$  & \cite{dellaValle1994} \\ 
     19 &GRS\,1009-45   & $3.8\pm0.3$  & $0.62\pm0.05$  & \cite{Gelino2002} \\ 
     17 &GRO\,J0422+32  & $2.5\pm0.3$  & $-0.54\pm0.06$ & \cite{Gelino2003} \\   
     15 &GS\,2023+338   & $2.4\pm0.1$  & $-0.08\pm0.01$ & \cite{Miller-Jones2009} \\ 
     13 &GS\,2000+251   & $2.7\pm0.7$  & $-0.14\pm0.04$ & \cite{Jonker2004} \\      
   \hline                           
     7 &A0620-003      & $1.1\pm0.1$  & $-0.11\pm0.01$ & \cite{Cantrell2010} \\
   \hline                           
  \end{tabular}
  }
\end{table}

There are 10 objects in this volume (listed in \tab\ref{tab:complete}). 
However, for the sake of the analysis we should only consider those BHTs detected since the time when the rate of discoveries has become more or less constant.
We assume this has occurred since 1988 (see \fig\ref{fig:histobhs}), when the sky started to be intensively scrutinized by all-sky monitors on-board X-ray satellites. Hence, we eliminate \acero~from this analysis because it was discovered in 1975, before the beginning of the assumed constant rate of discoveries. This yields 9 BHTs in the solar neighbourhood ($r\leq4$\,kpc), discovered at a rate of $\sim0.3$\,BHTs~yr$^{-1}$ since 1988. For comparison, in the same time interval, $\sim46$ BHTs have been discovered in the whole Galaxy (which implies a rate of $\simeq1.7$\,BHTs~yr$^{-1}$).

It is expected that the vertical distribution of BHTs follows the same exponential function as the stellar distribution \citep{Duerbeck1984}. In addition, it is reasonable to assume that the population of BHTs is representative of the parent Galactic population of BHTs since there is no strong bias against their detection in X-rays \citep{Ozel2010}.

Therefore, the space/time density distribution can be approximated by:
\begin{equation}
\label{equ:rhoz*}
    \rho^\ast(z) = \rho_0^\ast\,\exp\,\left(\frac{-|z|}{z_0}\right)\ ({\rm kpc}^{-3}\,{\rm yr}^{-1})
\end{equation}
where $\rho_0^\ast$ is the space/time density of objects in the Galactic plane and $z_0$ the scale height of the distribution perpendicular to the plane of the Galaxy, both in the solar neighbourhood.

\begin{figure}[!htp]
 \centering
    \includegraphics[width=0.5\txw,height=0.3\txh]{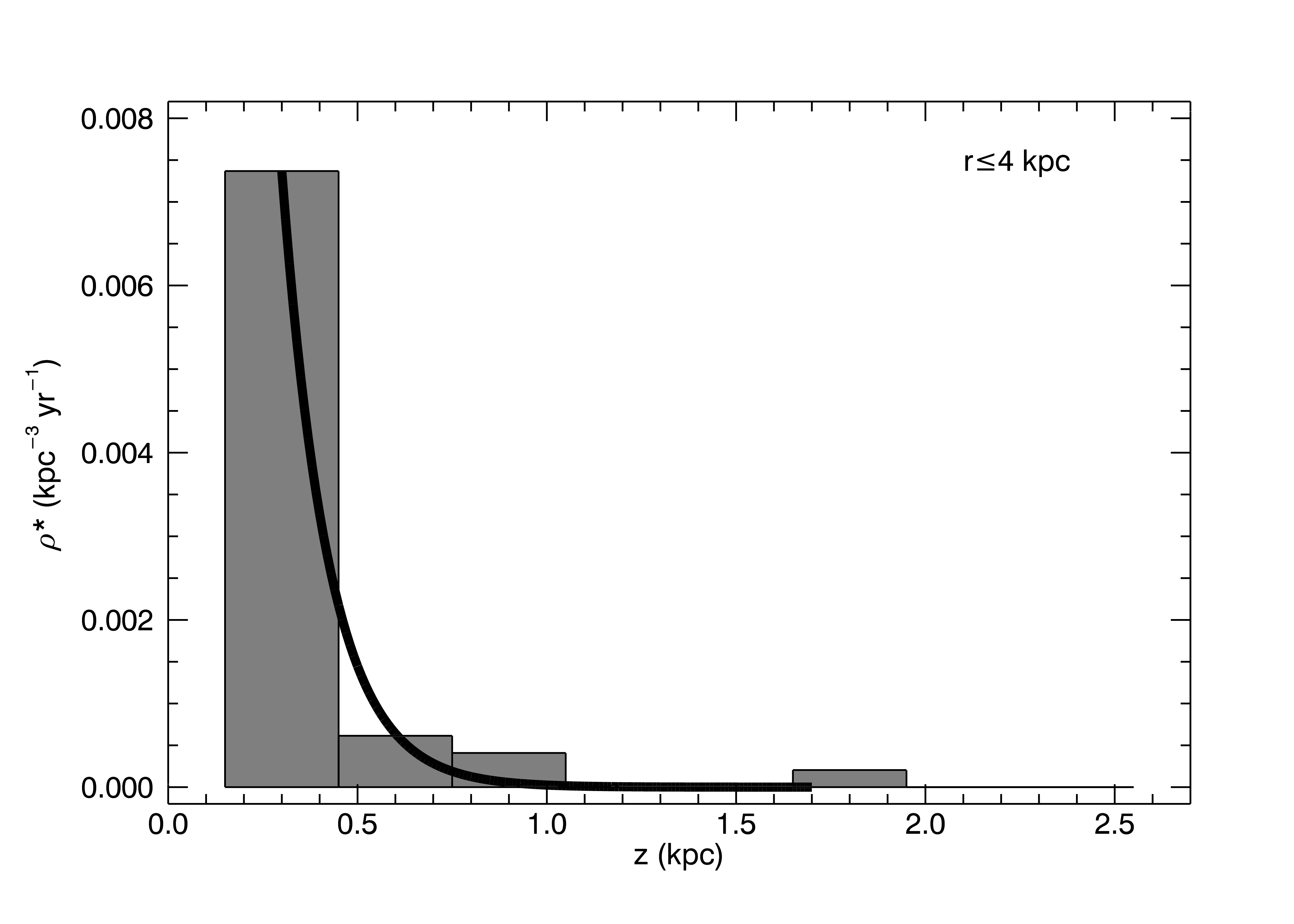}
    \caption[Variation of the space/time density of BHTs with the vertical distance]{Variation of the space/time density ($\rho^\ast$) of BHTs with the scale above the Galactic plane $z$. The solid line represents the Levenberg-Marquardt least-squares fit \citep{Markwardt2009} to the exponential function shown in \equ\ref{equ:rhoz*}.}
    \label{fig:spacetime}
\end{figure}

Using a Levenberg-Marquardt non-linear least-squares fit \citep{Markwardt2009} of $\rho(z)$, we derive $\rho^\ast_0=0.08\pm0.01$\,kpc$^{-3}$\,yr$^{-1}$ and $z_0=0.123\pm0.008$\,kpc (see \fig\ref{fig:spacetime}). This small scale height $z_0$ indicates a clear concentration of systems in the plane. 

\begin{figure}[!htp]
 \centering
    \includegraphics[width=0.5\txw]{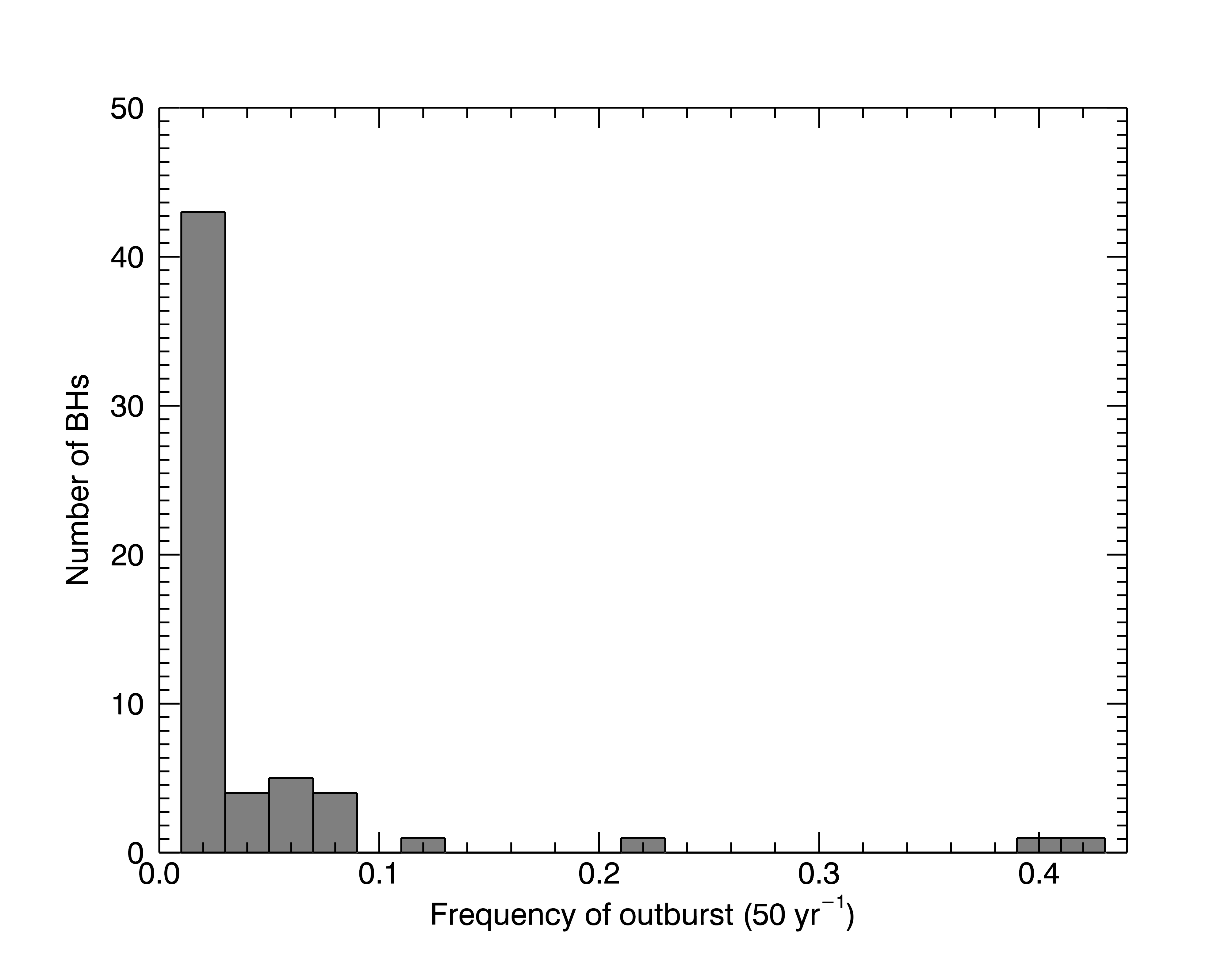}
    \caption[Frequency of outbursts detected in 50 years]
        {Histogram of the frequency of outbursts detected over 50\,yr period for all BHTs.}
    \label{fig:out_recur}
\end{figure}

To obtain the final space density of BHTs $\rho(z)$, we must assume a mean outburst recurrence period (ORP). We have only observed transients for $\sim50$ years and the observed ORPs are thus biased by this limited time-frame for scrutinizing the X-ray sky. In \fig\ref{fig:out_recur} we show the frequency of outbursts of all the BHTs over the past 50 years. The majority of systems have shown only one outburst, with only a small fraction showing multiple outbursts. The most extreme cases are H\,1743-322, GX\,339-4 and 4U\,1630-472, which have triggered 10, $\sim19$ and $\sim20$ outbursts in $\simeq50$ years (some of them were not considered ``full outbursts''), respectively. In particular, 4U\,1630-472 shows an ORP of 600-700\,d lasting from 100-200\,d up to 2.4 years (see, e.g., \citealt{Kuulkers1997}).
 There is also evidence for additional outbursts produced by A\,0620+003 \citep[in 1917; ][]{Eachus1976} and V404\,Cyg \citep[in 1938 and 1956; ][]{Wachmann1948,Richter1989} detected by analysing photographic plates previously to their discoveries. Nevertheless, in order to be consistent we included only those outbursts discovered by X-ray satellites (i.e. since 1966) and consequently, the observed ORP is biased by the $\simeq50$\,yr time lapse of X-ray astronomy observations. On the other hand, based on an analysis of the mass transfer rate needed to reach the critical surface density that produces instabilities in the accretion disc, \cite{White1996} consider unlikely that BHTs have average ORPs above 100\,yr. 
Therefore, the local density at the Galactic plane becomes $\rho_0 = {\rm ORP}\,\rho_0^\ast = 8\pm1 \left( \rm{\frac{ORP}{100\,yr}} \right)$\,kpc$^{-3}$.

A rough extrapolation of this local density distribution to the entire Galaxy, assuming no radial dependence, allows us to obtain the total number of BHTs by integrating \equ\ref{equ:rhoz*} in cylindrical coordinates.
Using $R=14$\,kpc as the truncation radius for the Galactic disc and $H=2.5$\,kpc as the maximum height above the Galactic plane given by \maxi~\citep{Kuulkers2013}, we find a total number of $N=1280\pm120 \left(\rm{\frac{ORP}{100\,yr}}\right)$ BHTs in the Galaxy with comparable properties to the systems detected so far, i.e. with peak X-ray luminosities a few tenths of the Eddington luminosity.
This is consistent with the results obtained using other techniques \citep{vandenHeuvel1992,Tanaka1992b,White1996,Romani1998} 
and implies that only $\sim4\%$ of Galactic BHTs have been discovered. 

Our empirical estimate is an order of magnitude lower than the $10^4$ BHTs predicted by \cite{Kiel2006} or \cite{Yungelson2006} using population synthesis models. However, we note that our analysis is based on the study of observed systems but limited only to the 9 BHTs with reliable distance estimates, located in a cylinder of 4\,kpc radius centred on the Sun. In addition, we have assumed that the solar vertical distribution ($\rho_0,z_0$) can be extrapolated to other regions of the Galaxy. However, the bulge contains $\sim30\%$ of the stellar mass of the Galaxy confined in a reduced spheroid and it is expected to host a higher concentration of BHTs \citep{Muno2005}. Furthermore, we have considered a cylinder with a height defined by \maxi~(the object with the highest $h$) but there could be objects located at higher distances over the plane. Finally, we have normalized our estimated value to an average recurrence period of 100\,yr, which does not consider those systems with smaller accretion rates and longer recurrence periods as well as a likely population of intrinsically faint X-ray BHTs. With all in hand, we conclude that our crude calculation of the number of BHTs expected in the Galaxy is very conservative and sets a lower limit to the hidden population.

\section{Physical properties of dynamical BHTs}
\label{sec:dynamical}

%
%
\begin{figure*}[!htp]
 \centering
    \includegraphics[width=0.45\txw]{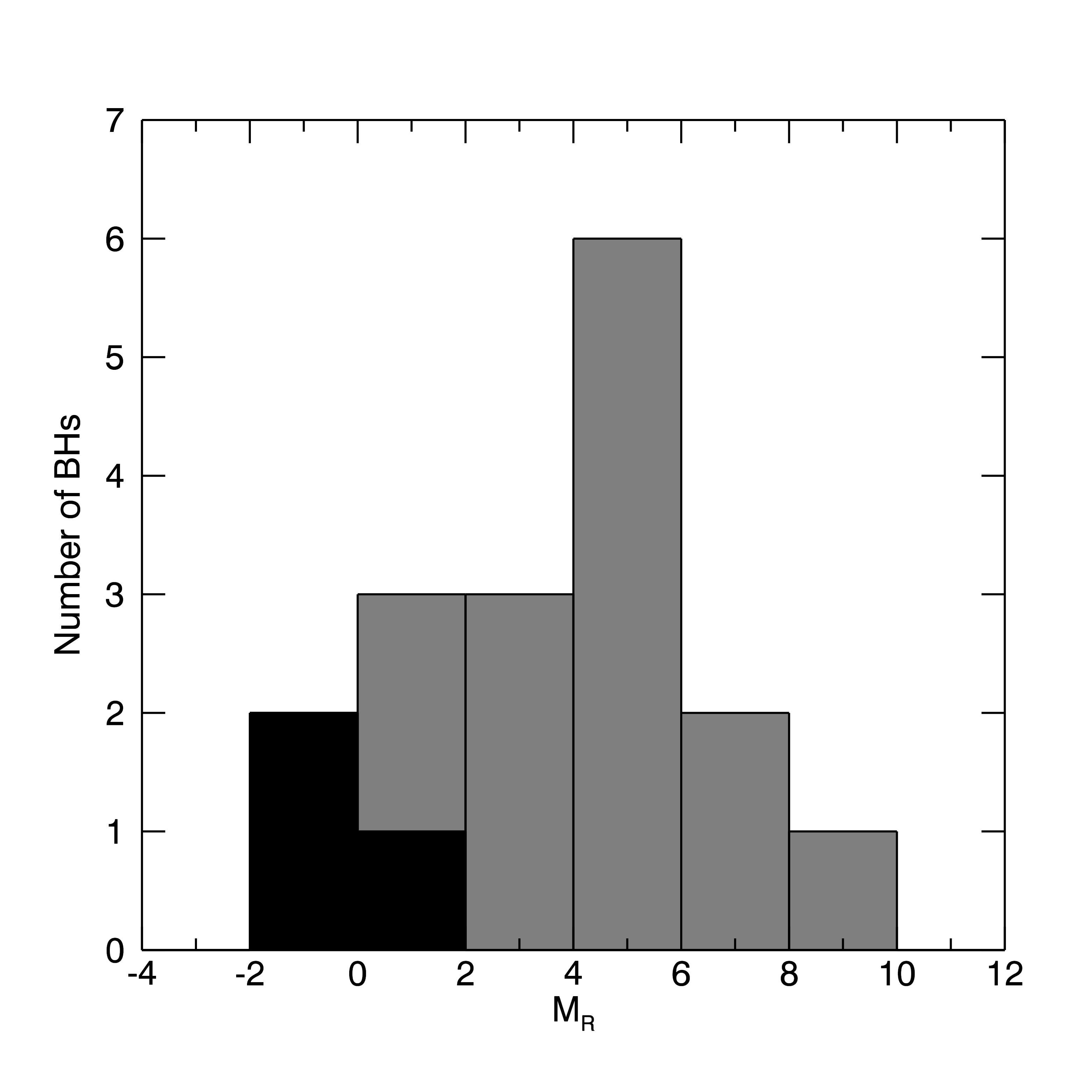}
    \includegraphics[width=0.45\txw]{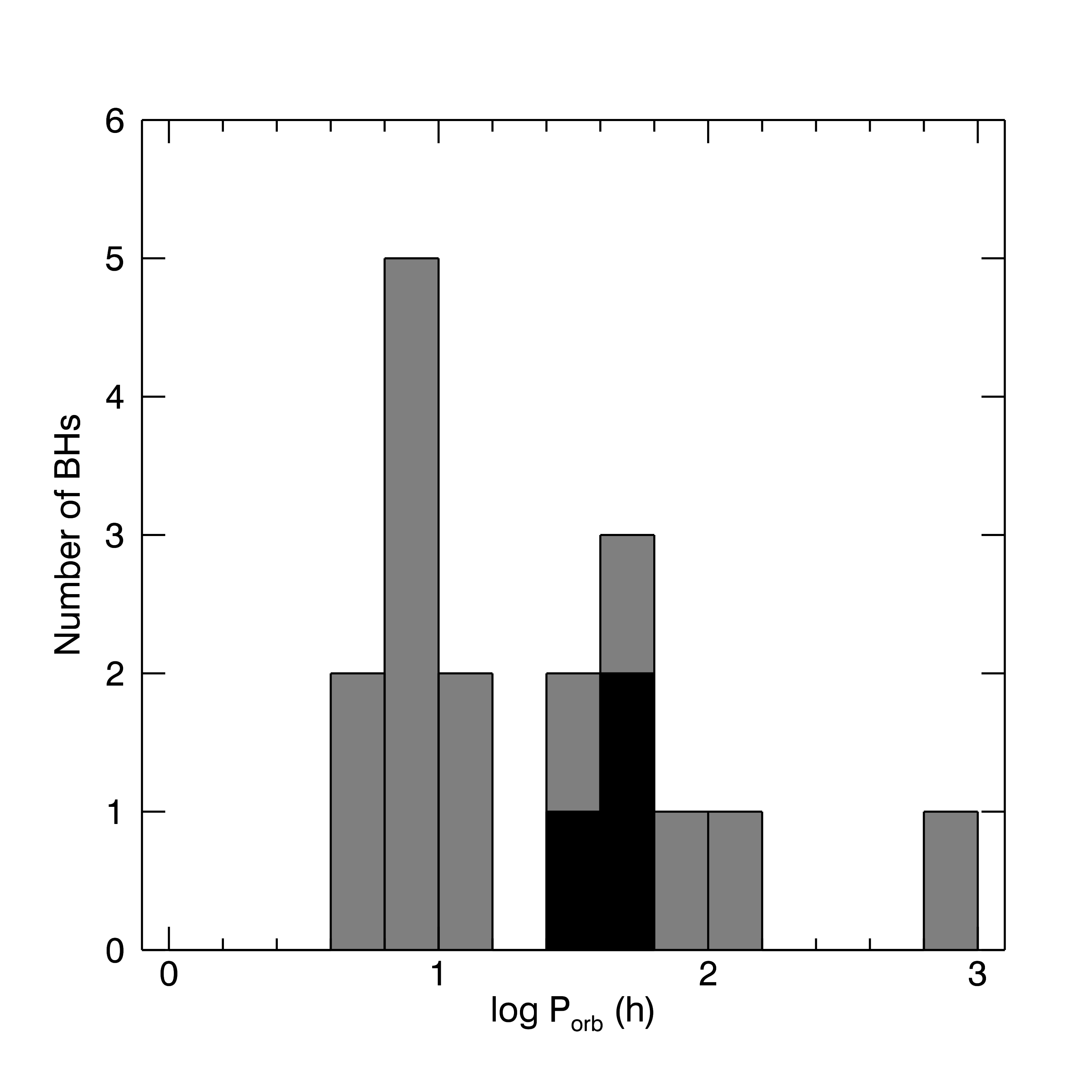}
    \caption[Histogram of orbital periods and absolute magnitudes]
        {Histograms of the 17 dynamically confirmed BHs. \textit{Left}: Extinction corrected absolute $R$ band magnitudes in bins of 2\,mag. The black histogram denotes confirmed IMXBs. \textit{Right}: Orbital periods in a logarithmic scale using bins of 0.2.}
    \label{fig:histo_p_m}
\end{figure*}

With the information on the apparent quiescent magnitudes, distances and reddening listed in \tab\ref{tab:census1}--\ref{tab:census4}, we can recover the $R$-band absolute magnitudes (corrected for extinction and in the AB system) of the BHTs in quiescence. The magnitude distribution (\fig\ref{fig:histo_p_m}, left) strongly peaks at $M_{\rm R}\simeq 4-6$, where $\sim40\%$ of the systems lie. This result is expected since the quiescent spectra are mainly dominated by the light from the donor (mostly K-type stars) with some contribution from the accretion flow (typically $\leq50$\% in this waveband). From \tab\ref{tab:census4}, we note a small fraction of IMXBs with A--F spectral type companions (V4641\,Sgr, 4U\,J1543--475 and GRO\,J1655--40), which have been marked in black in \fig\ref{fig:histo_p_m}.

In addition, from \tab\ref{tab:census2} we can revisit the observed distribution of orbital periods, which shows a bimodal shape with a gap at $\sim15$\,h known as the bifurcation period (\fig\ref{fig:histo_p_m}, right).
This is driven by different evolutionary paths, with systems above the gap evolving towards longer orbital periods through nuclear evolution of the donor stars, while angular momentum loses shrink the orbit of the binaries below the gap \citep{Pylyser1988,Menou1999}. In this case, the IMXBs are located at long orbital periods, consistent with the presence of giant/subgiant donor stars. The majority of systems have periods between 6--10\,h, which corresponds to main-sequence or slightly evolved K-type donors filling their Roche lobes.\\

It is noticeable that none of the 59 BHTs listed in \tab\ref{tab:census1}--\ref{tab:census2} present X-ray or optical eclipses (the extragalactic BH-HMXB M33~X-7 is the only one known to show eclipses). In fact, all the dynamical BHs transients found so far have binary inclinations $\lesssim75$\deg~(see \tab\ref{tab:census4}). 
However, it is expected that at least 20\% of the 17 dynamically confirmed BHs should have $i>75$\deg~considering an isotropic distribution of inclinations. Therefore, our estimate of the expected distribution of BHTs would be underestimated by 20\% and it should be $N\sim1600 \left(\frac{\rm ORP}{\rm 100\,yr}\right)$, 
although this percentage is probably smaller than the uncertainty associated to our systematic errors. \cite{Narayan2005a} have proposed that the lack of high inclination systems is due to a selection effect produced by the fact that the inner accretion disc hides or obscures the central BH, making these systems very faint in X-rays and preventing their detection during outbursts. For typical disc flaring angles of $\sim12$\deg~\citep{deJong1996}, the outer disc rim will obscure the central X-ray source permanently when viewed at inclinations $\gtrsim78$\deg. 

\jtrece~may be the first object seen edge-on according to the optical properties displayed during the decay from its 2011 outburst \citep{Corral-Santana2013}, but it did not show X-ray nor optical eclipses. 
\cite{ArmasPadilla2014a} and \cite{Torres2015} presented alternative explanations to the high inclination based on the quiescent X-ray properties and the Hydrogen column density obtained from the Na doublet in outburst, respectively. However, \cite{MataSanchez2015} presented new evidences supporting the edge-on configuration.

\begin{figure}[!ht]
 \centering
    \includegraphics[width=0.5\txw]{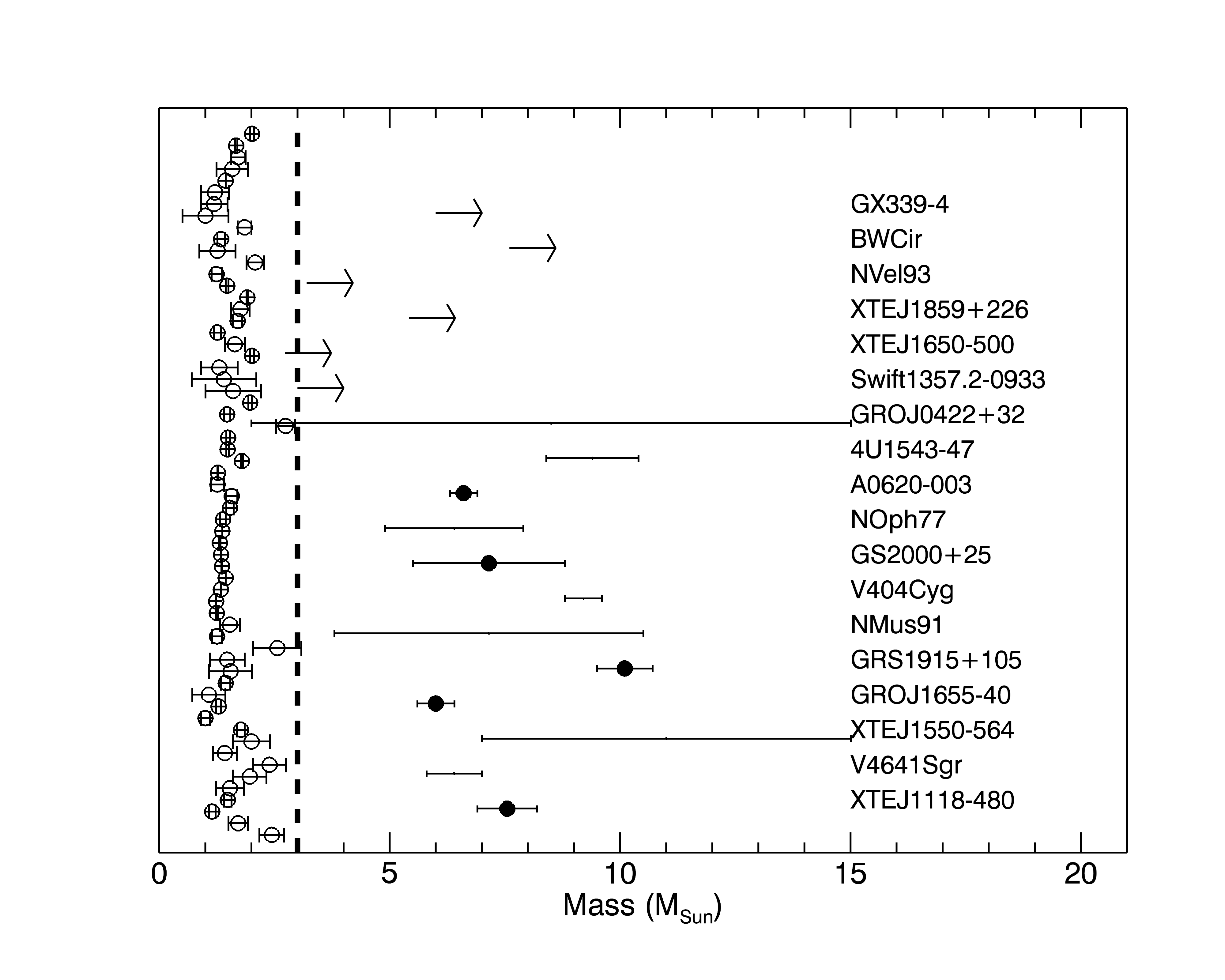}
    \caption{Distribution of observed compact object masses. The vertical dashed line represents the maximum mass allowed for NS \citep{Fryer2001}. Open circles below that limit represent the masses of the NS compiled by \cite{Lattimer2005}, extended with updated data from \cite{Ozel2012} and \cite{Antoniadis2013}. The solid circles indicate reliable BH masses (adopting the values favoured by \citealt{Casares2014a}) while arrows mark lower limits based on mass functions and upper limits to the inclination.
    }
    \label{fig:histo_mass}
\end{figure}

An accurate determination of the binary inclination is crucial for obtaining the mass of the compact object.
The mass distribution of black holes has a large impact on the physics of supernova explosions, the survival of interacting close binaries and the equation of state of nuclear matter. The distribution of masses of compact objects is expected to be smooth because of its correlation with the distribution of their progenitor masses \citep{Fryer2001}. However, the observed distribution shows a gap between neutron stars and BHs between $2$--$5$\msun~(\fig\ref{fig:histo_mass}). Unlike \cite{Ozel2010,Ozel2012}, here we only mark with solid circles those systems with the most reliable mass determinations following \cite{Casares2014a}. On the contrary, for those systems with inaccurate masses, we prefer to display large bars which encompass both the minimum and maximum published masses.

The existence of this mass gap is under debate. 
\cite{Ozel2010} argue that selection effects may bias the observed distribution at high BH masses ($\geq$10\msun) but can not explain the gap, which may be related to the physics of supernova explosions. \cite{Farr2011} perform a Bayesian analysis of the BH mass distribution and conclude that their larger sample provides strong evidence for the existence of the gap, reinforcing the results by \cite{Bailyn1998} and \cite{Ozel2010}.
However, \cite{Kreidberg2012} warn about an important source of systematic errors arising from a possible underestimate of the inclination angle $i$. This is produced by the assumption that the emission from the accretion disc (i.e., the disc veiling) is negligible in the infrared, 
when performing ellipsoidal modelling.
They found that in the case of \acero, this would lead to an underestimate of the inclination angle of 10\deg~at least. Because of the cubic dependence of the mass function with $\sin i$, this could lead to a considerable overestimate of the BH mass.
Correcting the BH mass in \jcero~from the estimated bias in the inclination, \cite{Kreidberg2012} find that the BH 
would be lie in the gap.
However, they also note that if this object is excluded from the analysis, previous conclusions would remain intact.\\
On the other hand, \cite{Belczynski2012} and \cite{Fryer2012} state that the mass gap may be real and it could reveal new insights on the supernovae explosion models.
More recently, \cite{Kochanek2014} suggested an alternative explanation based on the absence of red supergiants in the range $16.5-25$\msun~as progenitors of type IIp supernova. Due to the weakly bound hydrogen envelopes on these massive stars, they eject the outer layers leaving a BH with the mass of the star's helium core ($5-8$\msun). This would explain (i) the lack of supernova progenitors in the $16-25$\msun~mass range and (ii) the existence of the mass gap and the typical masses of BHs.

%
\section{Conclusions}
We have presented the main properties of a large catalogue of BHTs available on-line. Of the 59 BHTs detected in outburst so far, only $\sim30\%$ have been confirmed as dynamical BHTs. A fraction of them (35) have distance estimates which allow us to study the Galactic distribution. This in turn results in a population of $\sim1280 \left(\rm{\frac{ORP}{100\,yr}}\right)$ such systems in the Milky Way. This value is in agreement with previous estimates using other techniques but an order of magnitude lower than theoretical predictions based on population synthesis models. We argue that this value must be considered as a lower limit since it is based on the extrapolation of a small number of systems (9 out of the 59 systems detected so far).\\
We provide several tables listing the astrometric parameters, distances, number of eruptions, X-ray fluxes at the peak of the outburst and an outburst and a quiescence magnitude, reddening and orbital periods. For dynamically confirmed systems, we have also detailed the magnitudes in quiescence in every available optical and NIR band, together with their dynamical parameters. In the online version of the catalogue, we have also added finding charts, links to the references and relevant information for forthcoming observations of these systems. We plan to include more information in other wavelengths in the near future (e.g. radio or X-ray states) and update it with new targets once discovered.

\begin{longtab}
 \renewcommand{\arraystretch}{1.15}
\begin{landscape}
\scriptsize{
\centering
\begin{longtable}{lll lll ll ll l}
\caption[A census of BHTs (I): Astrometry]{\textbf{Astrometric properties}\\
         Columns: (1) Year of detection; (2) Name (and counterpart name); (3--5) RA--DEC coordinates (J2000), error in astrometry and source of the coordinates$\dagger$; 
         (6--7) Galactic longitude and latitude, respectively; (8--9) distance from Sun and above the Galactic plane; (10) number of outbursts reported after the discovery outburst and (11) references for detection, coordinates and distance.\\
}
\label{tab:census1}
\\
\hline
(1)  & (2)  & (3) & (4) & (5)    & (6) & (7) & (8) & (9)   & (10) & (11)\\
YEAR &NAME  & RA  & DEC & error$\dagger$ & $\ell$ & $b$ & $d$ & $z$ & Outb. & Ref. 
\\
     &      & (h~m~s) & (\deg~\arcmin~\arcsec) & (\arcsec)& (\deg) & (\deg) & (kpc) & (kpc) & &
\\
\hline
\hline
\hline
\endfirsthead
 
\multicolumn{9}{c}{\textbf{{\tablename} \thetable{}} -- Continued} \\
\hline
(1)  & (2)  & (3) & (4) & (5)    & (6) & (7) & (8) & (9)   & (10) & (11)\\
YEAR &NAME  & RA  & DEC & error$\dagger$ & $\ell$ & $b$ & $d$ & $z$ & Outb. & Ref. 
\\
     &      & (h~m~s) & (\deg~\arcmin~\arcsec) & (\arcsec)& (\deg) & (\deg) & (kpc) & (kpc) & &
\\
\hline
\hline
\hline
\endhead
  \\ \multicolumn{3}{l}{{Continued on next page\ldots}} \\
\endfoot

  \\[-1.8ex] \hline \hline
\endlastfoot
 
 & IGR J17454-2919\tablefootmark{1} & 17 45 27.69 & -29 19 53.83 & x 0.6 & 359.6444 & -00.1765 & $    $ & $    $ &  & \cite{Chenevez2014,Paizis2015} \\

\multirow{-2}{*}{2014}
 & IGR J17451-3022\tablefootmark{2} & 17 45 06.72 & -30 22 43.30 & x 0.6 & 358.7115 & -00.6580 & $    $ & $    $ &  & \cite{Chenevez2014b,Chakrabarty2014} \\

\hline
 & MAXI J1828-249 & 18 28 58.24 & -25 01 49.30 & o 0.03 & 008.1145 & -06.5458 & $    $ & $    $ &  & \cite{Nakahira2013,Kennea2013a} \\

\multirow{-2}{*}{2013}
 & SWIFT J1753.7-2544 & 17 53 39.85 & -25 45 14.20 & i 0.3 & 003.6476 & +00.1036 & $    $ & $    $ &  & \cite{Krimm2013,Rau2013} \\

\hline
 & SWIFT J174510.8-262411 & 17 45 10.85 & -26 24 12.60 & r (0.001,0.01) & 002.1107 & +01.4034  & $ <7^\ast$ & $<0.17$ &  & \cite{Cummings2012,Miller-Jones2012} 
 \\
 & & & & & & & & & & \cite{Munoz-Darias2013}
\\
 & SWIFT J1910.2-0546 & 19 10 22.80 & -05 47 55.92 & o 0.3& 029.9026 & -06.8440 &  $    $ & $    $ &  & \cite{Krimm2012,Usui2012,Rau2012} \\
 & (MAXI J1910-057) & & & & & & & & \\
\multirow{-5}{*}{2012}
 & MAXI J1305-704 & 13 06 55.30 & -70 27 05.11 & r (0.003,0.07)& 304.2375 & -07.6177 & $    $ & $    $ &  & \cite{Sato2012a,Coriat2012} 
\\
\hline
 & MAXI J1836-194\tablefootmark{3} & 18 35 43.44 & -19 19 10.48 & e (0.000003
,0.0002) & 013.9456 & -05.3542 & $ 7\pm 3$ & $-0.70\pm 0.30$ &  & \cite{Negoro2011b,Russell2015} \\

 & MAXI J1543-564\tablefootmark{4} & 15 43 17.18 & -56 24 49.61 & r (0.049,0.775) & 325.0855  & -01.1214 & $    $ & $    $ &  & \cite{Negoro2011a,Miller-Jones2011b} \\

\rowcolor{lightgray} \cellcolor{white}\multirow{-3}{*}{2011}
 & \textbf{SWIFT J1357.2-0933\tablefootmark{5}} & \textbf{13 57 16.82} & \textbf{-09 32 38.55} & \textbf{oi 0.3} & \textbf{328.7019} & \textbf{+50.0042} & $\mathbf{>2.29}$ & $\mathbf{>1.75}$ & \textbf{} & \textbf{\cite{Krimm2011a,Rau2011,MataSanchez2015}} \\

\hline
\multirow{-1}{*}{2010}
 & MAXI J1659-152 & 16 59 01.68 & -15 15 28.73 & e 0.0001 & 005.5003 & +16.5167 & $ 8.60\pm 3.70$ & $ 2.44\pm 1.05$ &  & \cite{Negoro2010,Paragi2010,Kuulkers2013} \\

\hline
 & XTE J1752-223 & 17 52 15.09 & -22 20 32.36 & e 0.0014 & 006.4231 & +02.1143 & $ 6\pm 2$ & $ 0.22\pm 0.07$ &  & \cite{Markwardt2009c,Miller-Jones2011,Ratti2012} \\

\multirow{-2}{*}{2009}
 & XTE J1652-453\tablefootmark{6} & 16 52 20.33 & -45 20 39.99 & r 0.32 & 340.5297 & -00.7867 & $    $ & $    $ &  & \cite{Markwardt2009a,Calvelo2009a} \\

\hline
 & SWIFT J1539.2-6227 & 15 39 11.96 & -62 28 02.30 & ox 0.5 & 321.0186 & -05.6427 & $    $ & $    $ &  & \cite{Krimm2008b,Krimm2011c} \\

\multirow{-2}{*}{2008}
 & SWIFT J1842.5-1124 & 18 42 17.45 & -11 25 03.90 & xo 0.6  & 021.7105 & -03.1076 &$    $ & $    $ &  & \cite{Krimm2008,Markwardt2008} \\

\hline
 & SWIFT J174540.2-290005\tablefootmark{7} & 17 45 40.10 & -29 00 06.40 & x 3.6 & 359.9495 & -00.0431 & $    $ & $    $ &  & \cite{Kennea2006,Kennea2006a} \\

 & IGR J17497-2821\tablefootmark{8} & 17 49 38.04 & -28 21 17.50 & i 0.1 & 000.9531 & -00.4527 & $    $ & $    $ &  & \cite{Soldi2006,Torres2006} \\

\multirow{-3}{*}{2006}
 & XTE J1817-330 & 18 17 43.53 & -33 01 07.57 & r 0.2 & 359.8172 & -07.9954 & $ 5.5\pm 4.5^\ast$ & $-0.80\pm 0.60$ &  & \cite{Remillard2006a,Rupen2006,Sala2007} \\

\hline
 & XTE J1726-476 & 17 26 49.28 & -47 38 24.90 & o 0.3 & 342.2032 & -06.9231 & $    $ & $    $ &  & \cite{Levine2005b,Turler2005,Maitra2005} \\
 & (IGR J17269-4737) & & & & & & & & \\
 & XTE J1818-245 & 18 18 24.43 & -24 32 17.96 & r (0.2,0.4) & 007.4427 & -04.1914 & $ 3.55\pm 0.75$ & $-0.26\pm 0.05$ &  & \cite{Levine2005a,Rupen2005a,CadolleBel2009} \\

 & SWIFT J1753.5-0127\tablefootmark{9} & 17 53 28.29 & -01 27 06.22 & r 0.05 & 024.8951 & 012.1842 & $\sim 6\pm 2$ & $\sim 1.30\pm 0.40$ &  & \cite{Palmer2005,Fender2005} \\

\multirow{-5}{*}{2005}
 & IGR J17098-3628 & 17 09 45.93 & -36 27 57.30 & r (0.011,0.55) & 349.5539 & 002.0745 & $\sim10.5$ & $\sim0.38$ &  & \cite{Grebenev2005,Rupen2005} \\

\hline
 & IGR J17091-3624\tablefootmark{10} & 17 09 07.61 & -36 24 25.7 & r 0.1 & 349.5249 & +02.2128 & $    $ & $    $ & 2 & \cite{Kuulkers2003,Rodriguez2011a} \\
 & (SAX J1709.1-3624) & & & & & & & & \\
\multirow{-3}{*}{2003}
 & XTE J1720-318 & 17 19 58.99 & -31 45 01.11 & r 0.25 & 354.6237  & +03.1013 & $ 6.5\pm 3.5$ & $ 0.4\pm 0.2$ &  & \cite{Remillard2003,OBrien2003,Chaty2006} \\
\hline
\multirow{-1}{*}{2002}
 & XTE J1908+094 & 19 08 53.08 & +09 23 04.84 & r (0.001,0.01) & 043.2615 & +00.4377 & $ 6.5\pm 3.5$ & $ 0.05\pm 0.03$ & 1 & \cite{Woods2002,Miller-Jones2013b,Chaty2006b} \\
\hline
 & SAX J1711.6-3808 & 17 11 37.10 & -38 07 05.70 & x 3.2 & 348.4200 & +00.7880 & $    $ & $    $ &  & \cite{intZand2001,intZand2002} \\
\rowcolor{lightgray} \cellcolor{white}\multirow{-2}{*}{2001}
 & \textbf{XTE J1650-500} & \textbf{16 50 00.98} & \textbf{-49 57 43.60} & \textbf{x 0.6} & \textbf{336.7182} & \textbf{-03.4270} & $\mathbf{ 2.6\pm 0.7}$ & $\mathbf{-0.16\pm 0.04}$ & \textbf{} & \textbf{\cite{Remillard2001,Tomsick2004,Homan2006}} \\
\hline
\rowcolor{lightgray} \cellcolor{white}
 & \textbf{XTE J1118+480\tablefootmark{11}} & \textbf{11 18 10.79} & \textbf{+48 02 12.42} & \textbf{r (0.004,0.02)} & \textbf{157.6607} & \textbf{+62.3206} & $\mathbf{1.7\pm 0.1}$ & $\mathbf{1.52\pm 0.09}$ & \textbf{1} & \textbf{\cite{Remillard2000,Fender2001b}} \\
\rowcolor{lightgray} \cellcolor{white}\multirow{-2}{*}{2000} & \textbf{(KV Uma)} & & & & & & & & & \\
\hline
\rowcolor{lightgray} \cellcolor{white}  & \textbf{XTE J1859+226} & \textbf{18 58 41.58} & \textbf{+22 39 29.40} & \textbf{o 0.5} & \textbf{054.0461} & \textbf{+08.6076} & $\mathbf{12.5\pm 1.5}$ & $\mathbf{ 1.90\pm 0.20}$ & \textbf{} & \textbf{\cite{Wood1999,Garnavich1999}} 
\\
\rowcolor{lightgray} \cellcolor{white} & \textbf{(V406 Vul)} & & & & & & & & & 
\textbf{\cite{Corral-Santana2011}}
\\
\rowcolor{lightgray} \cellcolor{white}\multirow{-2}{*}{1999}
 & \textbf{SAX J1819.3-2525} & \textbf{18 19 21.58} & \textbf{-25 24 25.10} & \textbf{o 0.7}& \textbf{006.7740} & \textbf{-04.7891} & $\mathbf{ 6.2\pm 0.7}$ & $\mathbf{-0.52\pm 0.06}$ & \textbf{3} & \textbf{\cite{intZand1999,Samus1999}} \\
\rowcolor{lightgray} \cellcolor{white} & \textbf{(V4641 Sgr)} & & & & & & & & & \textbf{\cite{MacDonald2014}}\\
\hline
 & XTE J2012+381\tablefootmark{12} & 20 12 37.71 & +38 11 01.10 & o 0.35 & 075.3883 & +02.2471 & $    $ & $    $ &  & \cite{Remillard1998,Hynes1999} 
\\
 & XTE J1748-288\tablefootmark{13} & 17 48 05.06 & -28 28 25.80 & r 0.6 & 000.6756 & -00.2220 & $\gtrsim8$ & $\sim-0.03$ &  & \cite{Smith1998b,Hjellming1998b} 
\\
 & & & & & & & & &  & \cite{Strohmayer1998} 
\\
\rowcolor{lightgray} \cellcolor{white}
 & \textbf{XTE J1550-564\tablefootmark{14}} & \textbf{15 50 58.70} & \textbf{-56 28 35.20} & \textbf{r 0.3} & \textbf{325.8825} & \textbf{-01.8269} & $\mathbf{4.5\pm0.5}$ & $\mathbf{-0.14\pm 0.02}$ & \textbf{5} & \textbf{\cite{Smith1998a,Corbel2001,Orosz2011a}} 
\\
\rowcolor{lightgray} \cellcolor{white}\multirow{-5}{*}{1998} & \textbf{(V381 Nor)} & & & & & & & & & 
\\
\hline
 & XTE J1755-324\tablefootmark{15} & 17 55 28.60 & -32 28 39.00 & x 60 & 358.0393 & -03.6314 & $    $ & $    $ &  & \cite{Remillard1997,Remillard1997} 
 \\
\multirow{-2}{*}{1997}
 & GRS 1737-31\tablefootmark{16} & 17 40 09.00 & -31 02 24.00 & x 30 & 357.5880 & -00.0990 & $    $ & $    $ &  & \cite{Sunyaev1997,Ueda1997} 
 \\
\hline
\\

 & GRS 1739-278\tablefootmark{17} & 17 42 40.03 & -27 44 52.70 & r (0.02,0.3) & 000.6721 & +01.1758 & $7.25\pm 1.25^\ast$ & $ 0.15\pm 0.03$ & 1 & \cite{Paul1996,Hjellming1996,Greiner1996} 
 \\
\multirow{-2}{*}{1996}
 & XTE J1856+053\tablefootmark{18} & 18 56 42.92 & +05 18 34.30 & i $<1$ & 038.2690 & +01.2720 & $    $ & $    $ & 2 & \cite{Marshall1996,Torres2007a} 
 \\
\hline
 & GRS 1730-312\tablefootmark{19} & 17 33 32.00 & -31 12 16.00 & x 180 & 356.6877 & +01.0065 & $    $ & $    $ &  & \cite{Borozdin1994,Vargas1996} 
 \\
 & (KS 1730-312) & & & & & & & & 
 \\
\rowcolor{lightgray} \cellcolor{white}\multirow{-2}{*}{1994}
 & \textbf{GRO J1655-40\tablefootmark{20}} & \textbf{16 54 00.14} & \textbf{-39 50 44.90} & \textbf{o (0.015,0.2)} & \textbf{344.9819} & \textbf{+02.4560} & $\mathbf{ 3.2\pm 0.2}$ & $\mathbf{ 0.14\pm 0.01}$ & \textbf{2 – 4} & \textbf{\cite{Zhang1994,Bailyn1995}} 
\\
\rowcolor{lightgray} \cellcolor{white} & \textbf{(N. Sco 1994)} & & & & & & & & & \textbf{\cite{Hjellming1995}}
\\
\hline
 & GRS 1716-249 & 17 19 36.93 & -25 01 03.43 & r 0.5 & 000.1423 & +06.9909 & $ 2.4\pm 0.4$ & $ 0.29\pm 0.05$ &  & \cite{Ballet1993,Harmon1993a} 
 \\
 & (N. Oph 1993) & & & & & & & & & \cite{Mirabel1993,dellaValle1994} 
\\
 & (V2293 Oph) & & & & & & & & & 
\\
\rowcolor{lightgray} \cellcolor{white}\multirow{-3}{*}{1993}
 & \textbf{GRS 1009-45} & \textbf{10 13 35.60} & \textbf{-45 04 35.28} & \textbf{o}& \textbf{275.8773} & \textbf{+09.3439} & $\mathbf{ 3.8\pm 0.3}$ & $\mathbf{ 0.62\pm 0.05}$ & \textbf{2} & \textbf{\cite{Lapshov1993,Harmon1993b}} 
 \\
\rowcolor{lightgray} \cellcolor{white} & \textbf{(N. Vel 1993 = MM Vel)} & & & & & & & & & \textbf{\cite{dellaValle1993,Gelino2002}} 
\\
\hline
\rowcolor{lightgray} \cellcolor{white} & \textbf{GRS 1915+105\tablefootmark{21}} & \textbf{19 15 11.55} & \textbf{+10 56 44.80} & \textbf{e 0.001} & \textbf{045.3656} & \textbf{-00.2194} & $\mathbf{10.4\pm 1.3}$ & $\mathbf{-0.040\pm 0.005}$ & \textbf{} & \textbf{\cite{Castro-Tirado1992,Dhawan2000}} 
\\
\rowcolor{lightgray} \cellcolor{white} & \textbf{(V1487 Aql)} & & & & & & & & & \textbf{\cite{Steeghs2013}}
\\
\rowcolor{lightgray} \cellcolor{white}\multirow{-2}{*}{1992}
 & \textbf{GRO J0422+32} & \textbf{04 21 42.79} & \textbf{+32 54 27.10} & \textbf{r 0.2} & \textbf{165.8790} & \textbf{-11.9108} & $\mathbf{ 2.49\pm 0.30}$ & $\mathbf{-0.51\pm 0.06}$ & \textbf{} & \textbf{\cite{Paciesas1992,Shrader1994}} 
 \\
\rowcolor{lightgray} \cellcolor{white} & \textbf{(V518 Per)} & & & & & & & & & \textbf{\cite{Gelino2003}} 
\\
\hline
\rowcolor{lightgray} \cellcolor{white}
 & \textbf{GRS 1124-684} & \textbf{11 26 26.65} & \textbf{-68 40 32.83} & \textbf{o (0.015,0.17)} & \textbf{295.3005} & \textbf{-07.0726} & $\mathbf{ 5.9\pm 0.3}$ & $\mathbf{-0.73\pm 0.04}$ & \textbf{} & \textbf{\cite{Lund1991,Makino1991}} 
 \\
\rowcolor{lightgray} \cellcolor{white}& \textbf{(N. Mus 1991)} & & & & & & & & & \textbf{\cite{dellaValle1991a,Hynes2005}}
\\
\rowcolor{lightgray} \cellcolor{white}\multirow{-3}{*}{1991}& \textbf{(GU Mus)} & & & & & & & & & 
\\
\hline
\rowcolor{lightgray} \cellcolor{white}
 & \textbf{GS 2023+338\tablefootmark{22}} & \textbf{20 24 03.82} & \textbf{+33 52 01.90} & \textbf{e (0.000002,0.00005)} & \textbf{073.1188} & \textbf{-02.0914} & $\mathbf{2.39\pm 0.14}$ & $\mathbf{-0.09\pm 0.01}$ & \textbf{2} & \textbf{\cite{Makino1989,Miller-Jones2009}} \\
\rowcolor{lightgray} \cellcolor{white}\multirow{-2}{*}{1989} & \textbf{(V404 Cyg)} & & & & & & & & & \\
\hline
 & GS 1734-275 & 17 36 02.00 & -27 25 41.00 & x 7 & 000.1608 & +02.5906 & $    $ & $    $ &  & \cite{Makino1988a,Voges1999} \\
 & (GRO 1735-27) & & & & & & & & \\
 & (KS 1732-273) & & & & & & & & \\
\rowcolor{lightgray} \cellcolor{white}\multirow{-3}{*}{1988}
 & \textbf{GS 2000+251} & \textbf{20 02 49.58} & \textbf{+25 14 11.30} & \textbf{r (0.07,1)} & \textbf{063.3666} & \textbf{-02.9989} & $\mathbf{ 2.7\pm 0.7}$ & $\mathbf{-0.14\pm 0.04}$ & \textbf{} & \textbf{\cite{Makino1988,Okamura1988}} \\
\rowcolor{lightgray} \cellcolor{white} & \textbf{(QZ Vul)} & & & & & & & & & \textbf{\cite{Hjellming1988,Jonker2004}}\\
\hline
\rowcolor{lightgray} \cellcolor{white}
 & \textbf{GS 1354-64\tablefootmark{23}} & \textbf{13 58 09.70} & \textbf{-64 44 05.80} & \textbf{r 0.2} & \textbf{309.9774} & \textbf{-02.7797} & $\mathbf{\sim25}$ & $\mathbf{\sim-1.21}$ & \textbf{2} & \textbf{\cite{Makino1987,Brocksopp2001}} 
 \\
\rowcolor{lightgray} \cellcolor{white}\multirow{-2}{*}{1987} & \textbf{(BW Cir)} & & & & & & & & & 
\\
\hline
 & EXO 1846-031\tablefootmark{24} & 18 49 16.92 & -03 03 54.74 & x 11 & 029.9585 & -00.9177 & $\sim7$ & $\sim-0.11$ &  & \cite{Parmar1985,Parmar1993} \\
\multirow{-2}{*}{1985}
 & SLX 1746-331\tablefootmark{25} & 17 49 48.94 & -33 12 11.60 & i 0.10 & 356.8092 & -02.9736 & $    $ & $    $ & 2 & \cite{Skinner1990,Torres2007b} \\
\hline
\rowcolor{lightgray} \cellcolor{white} & \textbf{H 1705-250} & \textbf{17 08 15.52} & \textbf{-25 05 30.15} & \textbf{o 0.2} &\textbf{358.5874} & \textbf{+09.0569} & $\mathbf{8.6\pm 2.1}$ & $\mathbf{ 1.40\pm 0.30}$ & \textbf{} & \textbf{\cite{Kaluzienski1977,Yang2012}} \\
\rowcolor{lightgray} \cellcolor{white} & \textbf{(N. Oph 1977)} & & & & & & & & & \textbf{\cite{Griffiths1978,Jonker2004}} 
\\
\rowcolor{lightgray} \cellcolor{white} & \textbf{(V2107 Oph)} & & & & & & & & &
\\
 & H 1743-322\tablefootmark{26} & 17 46 15.596 & -32 14 00.860 & r (0.0006,0.002) & 357.2552 & -01.8330 & $\sim10$ & $\sim-0.32$ & 10 &  \cite{White1983,Miller-Jones2012a} \\
 & (XTE J1746-322) & & & & & & & & \\
\multirow{-6}{*}{1977}
 & (IGR J17464-3213) & & & & & & & & \\
\hline
\rowcolor{lightgray} \cellcolor{white}
 & \textbf{3A 0620-003\tablefootmark{27}} & \textbf{06 22 44.50} & \textbf{-00 20 44.72} & \textbf{r (0.012,0.1)} & \textbf{209.3382} & \textbf{-06.2225} & $\mathbf{ 1.06\pm 0.10}$ & $\mathbf{-0.11\pm 0.01}$ & \textbf{1} & \textbf{\cite{Elvis1975,Gallo2006,Cantrell2010}} 
 \\
\rowcolor{lightgray} \cellcolor{white} & \textbf{(N. Mon 1975)} & & & & & & & & & 
\\
\rowcolor{lightgray} \cellcolor{white}\multirow{-3}{*}{1975} & \textbf{(V616 Mon)} & & & & & & & & & 
\\
\hline
\multirow{2}{*}{1974}
 & 3A 1524-617\tablefootmark{28} & 15 28 16.89 & -61 52 58.40 & o 0.3 & 320.3191 & -04.4272 & & & 1 & \cite{Pounds1974,Holt1974,Zurita2015} \\
 & (KY~Tra) & & & & & & & & \\
\hline
\rowcolor{lightgray} \cellcolor{white}
 & \textbf{1H 1659-487\tablefootmark{29}} & \textbf{17 02 49.40} & \textbf{-48 47 23.40} & \textbf{r 0.3} & \textbf{338.9445} & \textbf{-04.3229} & $\mathbf{>6}$ & $\mathbf{>-0.45}$ & $\mathbf{\sim19}$ & \textbf{\cite{Markert1973,Gallo2004}} 
\\
\rowcolor{lightgray} \cellcolor{white}\multirow{-2}{*}{1972} & \textbf{(GX 339-4)} & & & & & & & & & 
\\
\hline
 & 4U 1755-338\tablefootmark{30} & 17 58 40.04 & -33 48 26.80 & o 1 & 357.2155 & -04.8724 & $ 6.5\pm 2.5$ & $-0.60\pm 0.20$ &  & \cite{Giacconi1972,Bradt1983} \\
 & (V4134 Sgr) & & & & & & & & & \cite{Angelini2003} \\
\rowcolor{lightgray} \cellcolor{white}\multirow{-2}{*}{1971}
 & \textbf{4U 1543-475\tablefootmark{31}} & \textbf{15 47 08.32} & \textbf{-47 40 10.80} & \textbf{o 0.25} & \textbf{330.9266} & \textbf{+05.3626} & $\mathbf{ 7.5\pm 0.5}$ & $\mathbf{ 0.70\pm 0.05}$ & \textbf{3} & \textbf{\cite{Giacconi1972,Matilsky1972}} \\
\rowcolor{lightgray} \cellcolor{white} & \textbf{(IL Lup)} & & & & & & & & & \textbf{\cite{Park2004,Jonker2004}} \\
\hline
\multirow{2}{*}{1969}
 & 4U 1630-472\tablefootmark{32} & 16 34 01.61 & -47 23 34.80 & r (0.02,0.3) & 336.9112 & +00.2503 & $    $ & $    $ & $\sim20$ & \cite{Giacconi1972,Priedhorsky1986,Buxton1998} \\
 & (Nor X-1) & & & & & & & & \\
 \hline
\multirow{1}{*}{1966}
 & Cen~X-2\tablefootmark{33} & 14 00 28.2 & -64 47 35.6 & x 7200 & 310.2000 & -02.9000 & $    $ & $    $ & & \cite{Harries1967,Francey1971} \\
\hline
\end{longtable}
\noindent

\begin{flushleft}
\tablefoot{\\
  \textbf{In this catalogue we have included all the BH candidates detected so far. Dynamical BHs are highlighted in grey boxes.}\\ 
  $\dagger$~In column 5, the letter preceding the error value indicate the source of the coordinates: radio (r), interferometry in radio (e), optical (o), infrared (i), X-rays (x) or a combination of them. In case of different errors in RA and DEC (respectively), both are shown in parenthesis.\\
  $^\ast$~These values are uncertain estimates.\\
\tablefoottext{1}{IGR J17454-2919: the nature of this transient source is still unclear. \cite{Paizis2015} has proposed that 2MASS J17452768-2919534 is the IR counterpart although it is 2.4” away from the X-ray position.}\\
\tablefoottext{2}{IGR J17451-3022 has shown periodicity in its light curve of $\sim$6.3 h \citep{Jaisawal2015} which would imply a >70\deg~ inclination. However the nature of the compact object is still unclear and it might be a magnetar \citep{Heinke2014a}.}\\
\tablefoottext{3}{MAXI J1836-194: There is a field star 0.37\arcsec~away from the position of the BH candidate \citep{Russell2014}.}\\
\tablefoottext{4}{MAXI J1543-564 has no clear counterpart in optical/IR. \cite{Rau2011a} reported 3 possible objects close or within the radio \citep{Miller-Jones2011b} and X-ray \citep{Kennea2011a} error circles.}\\
\tablefoottext{5}{Swift J1357.2-0933 in rigour is not a dynamical BHT because it lacks a detection of the secondary star but it has robust mass function determination \citep{Corral-Santana2013,MataSanchez2015}. The distance is still uncertain (see \citealt{Rau2011,Shahbaz2013,MataSanchez2015}}.\\
\tablefoottext{6}{XTE J1652-453 has no clear counterpart in IR and there are doubts between close objects.}\\
\tablefoottext{7}{Swift J174540.2-290005 is 1 pc from Sgr A* assuming an 8.5 kpc distance \citep{Kennea2006}. If this transient is an X-ray binary it would be a LMXB \citep{Wang2006}.}\\
\tablefoottext{8}{IGR J17497-2821 seems to be blended with a bright star SE of the target \citep{Torres2006}. Probably located in the Galactic bulge.}\\
\tablefoottext{10}{Swift J1753.5-0127 remains in a low/hard X-ray state since its outburst in 2005.}\\
\tablefoottext{11}{IGR J17091-3624 showed outbursts in 2007 \citep{Capitanio2008} and 2011 \citep{Krimm2011}. Re-examination of archival data showed that is was also active in 1994, 1996 and 2001 \citep{Revnivtsev2003a,intZand2003,Capitanio2006}.}\\
\tablefoottext{12}{XTE J1118+480 showed an outburst in 2005 \citep{Zurita2005}.}\\
\tablefoottext{13}{XTE J2012+381 is blended with a star at 1.1” \citep{Hynes1999}.}\\
\tablefoottext{14}{XTE J1748-288 showed hard spectrum. No counterpart has been found. An unresolved radio source was detected \citep{Hjellming1998b} but it was located 1' away  from the center of the 1' uncertainty RXTE position \citep{Strohmayer1998}.}\\
\tablefoottext{15}{XTE J1550-564 showed a second complete outburst in 2000 \citep{Smith2000a} and other three faint mini-outbursts in 2001, 2002 and 2003 \citep{Tomsick2001,Belloni2002,Dubath2003}.}\\
\tablefoottext{16}{XTEJ1755-324 has an error radius of 1' in the coordinates \citep{Remillard1997}.}\\
\tablefoottext{17}{GRS 1737-31 is supposed to be close to the Galactic center \citep{Ueda1997}.}\\
\tablefoottext{18}{GRS 1739-278 showed another outburst in 2014 \citep{Miller2014a}.}\\
\tablefoottext{19}{XTE J1856+053 has shown two consecutive outbursts with $\sim$70 days interval between them \citep{Sala2008}. This behaviour was seen in the 1996 and 2007 outbursts \citep{Levine2007,Krimm2007,Sala2008}. This also appeared in the the recent 2015 outburst \citep{Suzuki2015,Negoro2015}.}\\
\tablefoottext{20}{KS 1730-312 has an error radius in the astrometry of 3' \citep{Vargas1996}.}\\
\tablefoottext{21}{GRO J1655-40 (N. Sco 1994) has shown several outburst.}\\
\tablefoottext{22}{GRS J1915+105 has remained in outburst since its discovery in 1992 \citep{Castro-Tirado1992}. It is included here because it was detected as a transient source.}\\
\tablefoottext{23}{GS 2023+338 (V404 Cyg) showed two outbursts detected in plate scales (1938 and 1956) prior to its detection in X-rays in 1989 \citep{Wachmann1948,Richter1989}. It has shown another outburst in 2015 \citep{Barthelmy2015}.}\\
\tablefoottext{24}{GS 1354-64 (BW Cir) is associated with MX 1353-64 which showed an outburst in 1971-1972. It showed another outburst in 1997 \citep{Remillard1997a} and 2015 \citep{Miller2015}.}\\
\tablefoottext{25}{EXO 1846-031 does not have a known counterpart. The error radius in the coordinates is 11” \citep{Parmar1993}.}\\
\tablefoottext{26}{SLX 1746-331 is a galactic centre source and therefore suffer of large extinction. It showed outbursts in 2003 \citep{Markwardt2003b} and 2007 \citep{Markwardt2007}. It also showed some activity in 2011 \citep{Ozawa2011}.}\\
\tablefoottext{27}{H 1743-322 has shown several outbursts.}\\
\tablefoottext{28}{3A 0620-003 showed an outburst in 1917 detected in plate scales prior to its detection in X-rays in 1975 \citep{Eachus1976}.}\\
\tablefoottext{29}{KY TrA showed another outburst in 1990 \citep{Barret1992}.}\\
\tablefoottext{30}{1H J1659-487 (GX 339-4) is a quasi persistent source.}\\
\tablefoottext{31}{4U 1755-338 is a quasi persistent system. This system stayed in outburst for 25\,yr until 1996 \citep{Roberts1996,Angelini2003}.}\\
\tablefoottext{32}{4U 1543-475 have shown outbursts in 1984, 1992 and 2002 \citep{Kitamoto1984,Harmon1992a,Miller2002}.}\\
\tablefoottext{33}{4U 1630-472: It has shown quasi-periodic outbursts with recurrence times of $\sim$600-700\,d and durations of 100-200\,d \citep{Kuulkers1997}.}\\
\tablefoottext{34}{Cen X-2 is the first transient XRB ever discovered. It has an uncertain position. \cite{Kitamoto1990} proposed that it might be GS 1354-64 (BW Cir), rediscovered in 1987. However, if this were true, it would imply an extreme X-ray flux at its distance of 25 kpc \citep{Casares2009}.}\\
}
\end{flushleft}
}
\end{landscape}
\renewcommand{\thefootnote}{\arabic{footnote}}
\renewcommand{\arraystretch}{1.}

 \renewcommand{\arraystretch}{1.15}
\begin{landscape}
\scriptsize{
\centering
\begin{longtable}{lll l ll l l}
\caption[A census of BHTs (II): Photometric properties]{\textbf{Peak X-ray flux and optical/NIR photometric parameters}\\
Columns: (1) ID number used in this catalogue, (2) name (and counterpart name), (3) X-ray flux at the peak of the discovery outburst in the 2-10 keV band, (4--5) outburst and quiescent AB magnitudes (keeping the name of the band in the original system), (6) reddening, (7) orbital period and (8) references. 
} 
\label{tab:census2}
\\
\hline
(1) & (2)  & (3)                       & (4)   & (5)    & (6)         & (7)       & (8)\\
ID  & NAME & f$_x^{PEAK}$ [2-10\,keV]    & Outb. & Quies. & $E(B-V)^\S$ & P$_{\rm ORB}$ & Ref.\\
      &      & ($\rm erg s^{-1} cm^{-2}$)& (AB mag) & (AB mag)  & (mag)       & (h)       &\\
\hline
\hline
\endfirsthead
 
\multicolumn{7}{c}{{\textbf{\tablename~\thetable{}}} -- Continued} \\
\hline
(1) & (2) & (3) & (4) & (5) & (6) & (7) & (8)\\
ID  & NAME & f$_x^{PEAK}$ [2-10\,keV]     & Outb.    & Quies.    & $E(B-V)^\S$ & P$_{ORB}$ & Ref.\\
      &      & ($\rm erg s^{-1} cm^{-2}$) & (AB mag) & (AB mag)  & (mag)          & (h)       &\\
\hline
\hline
\endhead
 
  \multicolumn{3}{l}{{Continued on next page\ldots}} \\
\endfoot
 
  \\[-1.8ex] \hline \hline
\endlastfoot
59 & IGR J17454-2919 & \xe{2.78}{-10} & $ $ & $J=17.137$ & $>10^\S$ &$ $ &\cite{Chenevez2014a} 
\\ 
58 & IGR J17451-3022 & \xe{1.90}{-10} & $ $ & $ $ & $>10^\S$ &$\sim6.284\pm0.001$ &\cite{Heinke2014a,Altamirano2014,Jaisawal2015} 
\\
\hline
57 & MAXI J1828-249 & \xe{4.58}{-09} & $r'=16.9\pm0.1$ & $ $ & $0.34^\S$ &$ $ &\cite{Filippova2014,Rau2013a} 
\\
56 & SWIFT J1753.7-2544 & \xe{7.09}{-09} & $K\sim16.5^\ast$ & $ $ & $>10$ &$ $ &\cite{Krimm2013,Rau2013,Schlafly2011} 
\\
\hline
55 & SWIFT J174510.8-262411\tablefootmark{1} & \xe{2.90}{-08} & $i'\sim17.6$ & $r'>23.1\pm0.5$ & $2.87^\S$ &$\leq21$ &\cite{Munoz-Darias2013,Hynes2012a} 
\\
54 & SWIFT J1910.2-0546 & \xe{1.95}{-08} & $r'=15.7\pm0.1$ & $ $ & $0.52^\S$ &$>6.2$ &\cite{Reis2013,Rau2012,Casares2012} 
\\
& (MAXI J1910-057) & & & & & & \\
53 & MAXI J1305-704 & \xe{1.03}{-09} & $uvw1=17.23\pm0.04$ & $ $ & $0.29^\S$ &$9.74\pm0.04$ &\cite{Morihana2013,Greiner2012,Shidatsu2013} 
\\
\hline
52 & MAXI J1836-194 & \xe{1.01}{-09} & $V=16.33\pm0.08$ & $r>23.474$ & $0.6$ &$<4.9$ &\cite{Russell2013a} 
\\
51 & MAXI J1543-564 & \xe{1.43}{-09} & $ $ & $ $ & $3.518$ &$ $ &\cite{Stiele2012,Kennea2011c} 
\\
\rowcolor{lightgray}\cellcolor{white}50 & \textbf{SWIFT J1357.2-0933} & \xebf{2.23}{-10} & $\mathbf{r'=16.3\pm0.05}$ & $\mathbf{r'=21.5\pm0.4}$ & $\mathbf{0.037^\S}$ &$\mathbf{2.8\pm0.3}$ &\textbf{\cite{ArmasPadilla2013,Rau2011}} 
\\
\rowcolor{lightgray}\cellcolor{white} &&&&&&&\textbf{\cite{Shahbaz2013,Corral-Santana2013}}
\\
\hline
49 & MAXI J1659-152 & \xe{6.80}{-09} & $V=16.756$ & $r'\gtrsim23.7\pm0.1$ & $0.34$ &$2.414\pm0.005$ &\cite{Munoz-Darias2011,Russell2010,Kong2012} 
\\
 &&&&&&&\cite{DAvanzo2010,Kuulkers2013}
\\
\hline
48 & XTE J1752-223 & \xe{5.18}{-10} & $J=15.75\pm0.01$ & $i'>24.4$ & $1.403^\S$ &$<7$ &\cite{Stiele2011,Torres2009a,Ratti2012} 
\\
47 & XTE J1652-453 & \xe{1.39}{-08} & $Ks=18.8\pm0.3$ & $ $ & $8.799^\S$ &$ $ &\cite{Markwardt2009b,Torres2009b} 
\\
\hline
46 & SWIFT J1539.2-6227\tablefootmark{2} & \xe{6.47}{-09} & $uvw2=18.07\pm0.03$ & $ $ & $0.386^\S$ &$ $ &\cite{Krimm2011c} 
\\
45 & SWIFT J1842.5-1124 & \xe{9.85}{-09} & $Ks=16.75\pm0.05$ & $Ks\sim19.25$ & $0.767^\S$ &$ $ &\cite{Krimm2008c,Torres2008} 
\\
\hline
44 & SWIFT J174540.2-290005 & \xe{4.98}{-12} & $Ks>19.55$ & $ $ & $>10^\S$ &$ $ &\cite{Kennea2006,Wang2006} 
\\
43 & IGR J17497-2821 & \xe{1.09}{-09} & $Ks=17.8\pm0.2$ & $ $ & $>10^\S$ &$ $ &\cite{Paizis2007,Torres2006} 
\\
42
 & XTE J1817-330 & \xe{4.94}{-08} & $g'=14.93\pm0.05$ & $V>21.956$ & $0.214^\S$ &$ $ &\cite{Sala2007,Torres2006a} 
\\
\hline
41 & XTE J1726-476 & \xe{3.53}{-09} $\dagger$ & $I=17.3\pm0.1$ & $V>18.956$ & $0.428^\S$ &$ $ &\cite{Steeghs2005b} 
\\
& (IGR J17269-4737) & & & & & & 
\\
40 & XTE J1818-245\tablefootmark{3} & \xe{1.69}{-08} & $R\sim17.555$ & $R=16.26\pm0.3$ & $0.758^\S$ &$ $ &\cite{CadolleBel2009,Steeghs2005} 
\\
39 & SWIFT J1753.5-0127\tablefootmark{4} & \xe{5.64}{-09} & $R\sim15.855$ & $V>20.956$ & $0.34$ &$2.85\pm0.01$ &\cite{Soleri2013,Halpern2005,CadolleBel2007,Neustroev2014} 
\\
38 & IGR J17098-3628 & \xe{1.08}{-09} & $V\sim20.756$ & $ $ & $\sim1.45$ &$ $ &\cite{Grebenev2007,Steeghs2005a,Steeghs2005c}
\\
\hline
37 & IGR J17091-3624\tablefootmark{5} & \xe{5.96}{-09} & $I=18.66\pm0.03$ & $ $ & $1.921^\S$ &$ $ &\cite{Altamirano2011,Torres2011} 
\\
& (SAX J1709.1-3624) & & & & & & 
\\
36 & XTE J1720-318\tablefootmark{6} & \xe{1.06}{-08} $\dagger$ & $Ks\sim17.15$ & $ $ & $\geq2.25$ &$ $ &\cite{Nagata2003,Chaty2006} 
\\
\hline
35 & XTE J1908+094\tablefootmark{7} & \xe{3.77}{-09} & $R>23$ & $J\sim21.1\pm0.1$ & $5.415^\S$ &$ $ &\cite{intZand2002b,Wagner2002} 
\\
\hline
34 & SAX J1711.6-3808 & \xe{1.27}{-09} & $ $ & $R>22.055$ & $5.161$ &$ $ &\cite{Wijnands2002,intZand2002,Wang2014} 
\\
\rowcolor{lightgray}\cellcolor{white}33 & \textbf{XTE J1650-500\tablefootmark{9}} & \xebf{1.52}{-08} & $\mathbf{B\sim16.837}$ & $\mathbf{R\sim22.055}$ & $\mathbf{1.5}$ &$\mathbf{7.69\pm0.02}$ &\textbf{\cite{Corbel2004,Castro-Tirado2001,Garcia2002}} 
\\
\rowcolor{lightgray}\cellcolor{white}&&&&&&& \textbf{\cite{Curran2012,Augusteijn2001a,Orosz2004}}
\\
\hline
\rowcolor{lightgray}\cellcolor{white}32 & \textbf{XTE J1118+480\tablefootmark{10}} & \xebf{4.99}{-10} & $\mathbf{V=13\pm0.3}$ & $\mathbf{R=19.055}$ & $\mathbf{0.024}$ &$\mathbf{4.078414\pm0.000005}$ &\textbf{\cite{Brocksopp2010,Torres2002,Gelino2006}} 
\\
\rowcolor{lightgray}\cellcolor{white} & \textbf{(KV Uma)} & & & & & & \textbf{\cite{Garcia2000,Torres2004}} 
\\
\hline
\rowcolor{lightgray}\cellcolor{white}31 & \textbf{XTE J1859+226\tablefootmark{11}} & \xebf{3.35}{-08} & $\mathbf{R\sim15.155}$ & $\mathbf{R=22.535\pm0.07}$ & $\mathbf{0.58}$ &$\mathbf{6.58\pm0.05}$ &\textbf{\cite{Wood1999,Garnavich1999,Zurita2002}} 
\\
\rowcolor{lightgray}\cellcolor{white}& \textbf{(V406 Vul)} & & & & & & \textbf{\cite{Hynes2002,Corral-Santana2011}}
\\
\rowcolor{lightgray}\cellcolor{white}30 & \textbf{SAX J1819.3-2525} & \xebf{2.76}{-07} & $\mathbf{V=8.756}$ & $\mathbf{I=13.519\pm0.03}$ & $\mathbf{0.32-0.38}$ &$\mathbf{67.6152\pm0.0002}$ &\textbf{\cite{Orosz2001,MacDonald2014}} 
\\
\rowcolor{lightgray}\cellcolor{white} & \textbf{(V4641 Sgr)} & & & & & & 
\\
\hline
29 & XTE J2012+381 & \xe{7.77}{-09} & $R=20\pm0.2$ & $ $ & $1.9-2.4$ &$ $ &\cite{Hynes1999} \\ 
28 & XTE J1748-288 & \xe{1.61}{-08} & $ $ & $ $ & $>10^\S$ &$ $ &\cite{Revnivtsev2000} 
\\
\rowcolor{lightgray}\cellcolor{white}27 & \textbf{XTE J1550-564} & \xebf{1.61}{-07} & $\mathbf{V\sim16.556}$ & $\mathbf{i=19.044\pm0.03}$ & $\mathbf{1.33}$ &$\mathbf{37.0088\pm0.0001}$ &\textbf{\cite{Wu2002,Orosz2002}} 
\\
\rowcolor{lightgray}\cellcolor{white}& \textbf{(V381 Nor)} & & & & & & \textbf{\cite{Calvelo2010,Orosz2011a}}
\\
\hline
26 & XTE J1755-324 & \xe{3.41}{-09} & $ $ & $ $ & $0.988^\S$ &$ $ &\cite{Revnivtsev1998} 
\\
25 & GRS 1737-31 & \xe{6.00}{-10} & $ $ & $ $ & $>10^\S$ &$ $ &\cite{Marshall1997} \\
\hline
\\
\\

24 & GRS 1739-278 & \xe{7.20}{-09}$\ddagger$ & $R=20.6\pm0.1$ & $J>19.2 $ & $2-4$ &$ $ &\cite{Marti1997,Chaty2002a} 
\\
23 & XTE J1856+053 & \xe{7.46}{-10} & $Ks=18.28\pm0.05$ & $ $ & $5.113^\S$ &$ $ &\cite{Sala2008,Torres2007a} 
\\
\hline
22 & GRS 1730-312\tablefootmark{13} & \xe{1.14}{-08} & $ $ & $ $ & $5.5^\S$ &$ $ &\cite{Borozdin1995} 
\\
& (KS 1730-312) & & & & & & 
\\
\rowcolor{lightgray}\cellcolor{white}21 & \textbf{GRO J1655-40\tablefootmark{14}} & \xebf{1.15}{-07} & $\mathbf{R=13.555}$ & $\mathbf{R=16.3\pm0.1}$ & $\mathbf{1.3}$ &$\mathbf{62.92\pm0.003}$ &\textbf{\cite{Brocksopp2006,Bailyn1995,Orosz1997}} 
\\
\rowcolor{lightgray}\cellcolor{white}& \textbf{(N. Sco 1994)} & & & & & & \textbf{\cite{Greene2001,Horne1996,VanDerHooft1998}}
\\
\hline
20 & GRS 1716-249 & \xe{1.97}{-08} & $V\sim16.606$ & $R\gtrsim20.055$ & $0.9$ &$ $ &\cite{Revnivtsev1998a,dellaValle1994} 
\\
& (N. Oph 1993 = V2293 Oph) & & & & & & 
\\
\rowcolor{lightgray}\cellcolor{white}19 & \textbf{GRS 1009-45\tablefootmark{15}} & \xebf{5.10}{-08} & $\mathbf{R<14.705}$ & $\mathbf{R=21.3\pm0.2}$ & $\mathbf{0.18-0.23}$ &$\mathbf{6.8449\pm0.0003}$ &\textbf{\cite{Borozdin1993a,dellaValle1997,Shahbaz1996}} 
\\
\rowcolor{lightgray}\cellcolor{white}& \textbf{(N. Vel 1993 = MM Vel)} & & & & & & \textbf{\cite{Filippenko1999,Hynes2003}} 
\\
\hline
\rowcolor{lightgray}\cellcolor{white}18 & \textbf{GRS 1915+105\tablefootmark{16}} & \xebf{9.29}{-08} $\dagger$ & $\mathbf{K=11.4^\ast}$ & $\mathbf{I\sim23.7\pm0.4}$ & $\mathbf{>>6.45}$ &$\mathbf{812\pm4}$ &\textbf{\cite{Shahbaz2008,Boeer1996}} 
\\
\rowcolor{lightgray}\cellcolor{white}& \textbf{(V1487 Aql)} & & & & & & 
\textbf{\cite{Fender1999,Steeghs2013}}
\\
\rowcolor{lightgray}\cellcolor{white}17 & \textbf{GRO J0422+32} & \xebf{5.77}{-08} & $\mathbf{R=12.655}$ & $\mathbf{R=21\pm0.1}$ & $\mathbf{0.3}$ &$\mathbf{5.09185\pm0.000005}$ &\textbf{\cite{Goldwurm1992,Castro-Tirado1992b,Gelino2003}} 
\\
\rowcolor{lightgray}\cellcolor{white}& \textbf{(V518 Per)} & & & & & & \textbf{\cite{Garcia1996,Jonker2004,Webb2000}} 
\\
\hline
\rowcolor{lightgray}\cellcolor{white}16 & \textbf{GRS 1124-684\tablefootmark{17}} & \xebf{1.36}{-07} & $\mathbf{V\sim13.456}$ & $\mathbf{i'=19.92\pm0.04}$ & $\mathbf{0.3}$ &$\mathbf{10.38254\pm0.00007}$ &\textbf{\cite{Brandt1992,dellaValle1991a}} 
\\
\rowcolor{lightgray}\cellcolor{white}& \textbf{(N. Mus 1991 = GU Mus)} & & & & & & \textbf{\cite{Shahbaz2010,Orosz1996}} 
\\
\hline
\rowcolor{lightgray}\cellcolor{white}15 & \textbf{GS 2023+338} & \xebf{2.59}{-07} & $\mathbf{V\sim11.7}$ & $\mathbf{R=16.58\pm0.01}$ & $\mathbf{1.3}$ &$\mathbf{155.311\pm0.002}$ &\textbf{\cite{Zycki1999,Zurita2004}} 
\\
\rowcolor{lightgray}\cellcolor{white} & \textbf{(V404 Cyg)} & & & & & & \textbf{\cite{Casares1991,Casares1992,Casares1993,Casares1994}}
\\
\hline
14 & GS 1734-275 & \xe{1.20}{-09} & $ $ & $ $ & $1.322^\S$ &$ $ &\cite{Yamauchi2004} \\
& (GRO 1735-27 = KS 1732-273) & & & & & & 
\\
\rowcolor{lightgray}\cellcolor{white}13 & \textbf{GS 2000+251} & \xebf{3.46}{-07} & $\mathbf{V=16.356}$ & $\mathbf{R=21.3\pm0.2}$ & $\mathbf{1.1-1.7}$ &$\mathbf{8.258095\pm0.000005}$ &\textbf{\cite{Tsunemi1989,Wagner1988,Charles1988,Charles1991}} 
\\
\rowcolor{lightgray}\cellcolor{white}& \textbf{(QZ Vul)} & & & & & & \textbf{\cite{Callanan1991,Casares1995a,Ioannou2004}} 
\\
\hline
\rowcolor{lightgray}\cellcolor{white}12 & \textbf{GS 1354-64} & \xebf{2.38}{-09} & $\mathbf{V=16.876}$ & $\mathbf{R=20.71\pm0.03}$ & $\mathbf{1}$ &$\mathbf{61.068\pm0.002}$ &\textbf{\cite{Kitamoto1990,Casares2009}} 
\\
\rowcolor{lightgray}\cellcolor{white}& \textbf{(BW Cir)} & & & & & & \\
\hline
11 & EXO 1846-031\tablefootmark{18} & \xe{1.02}{-08} & $ $ & $I>22.309$ & $6.437^\S$ &$ $ &\cite{Parmar1993} 
\\
10 & SLX 1746-331\tablefootmark{19} & \xe{1.37}{-08} $\dagger$ & $Ks=18.78\pm0.06$ & $ $ & $1.278^\S$ &$ $ &\cite{Torres2007b} 
\\
\hline
\rowcolor{lightgray}\cellcolor{white}9 & \textbf{H 1705-250} & \xebf{1.75}{-08} & $\mathbf{B=16.3\pm0.5}$ & $\mathbf{R\sim20.9\pm0.2}$ & $\mathbf{0.5}$ &$\mathbf{12.51\pm0.03}$ &\textbf{\cite{Watson1978,Longmore1977,Griffiths1978}} 
\\
\rowcolor{lightgray}\cellcolor{white}& \textbf{(N. Oph 1977 = V2107 Oph)} & & & & & & \textbf{\cite{Remillard1996,Martin1995}} 
\\
8 & H 1743-322\tablefootmark{20} & \xe{4.62}{-08} $\dagger$ & $R=21.955$ & $i'>24$ & $3.049$ &$ $ &\cite{Steeghs2003,Schlegel1998,McClintock2009} 
\\
& (XTE J1746-322 = IGR J17464-3213) & & & & & & 
\\
\hline
\rowcolor{lightgray}\cellcolor{white}7& \textbf{3A 0620-003} & & & & & & \textbf{\cite{Warwick1981,Robertson1976}} 
\\
\rowcolor{lightgray}\cellcolor{white} & \textbf{(N. Mon 1975)} & \xebf{2.65}{-06} & $\mathbf{V\sim11.12}$ & $\mathbf{R=17.12\pm0.04}$ & $\mathbf{0.35}$ &$\mathbf{7.7523372\pm0.0000002}$ & \textbf{\cite{Cantrell2010,Gelino2001}} 
\\
\rowcolor{lightgray}\cellcolor{white}& \textbf{(V616 Mon)} & & & & & &\textbf{\cite{Shahbaz1998,GonzalezHernandez2010}} 
\\
\hline
6 & 3A 1524-617\tablefootmark{21} & \xe{4.66}{-08} & $B=17.337$ & $R=22.4\pm0.1$ & $0.629^\S$ &$8-15$ &\cite{Kaluzienski1975,Murdin1977,Zurita2015} 
\\
& (KY Tra) & & & & & & 
\\
\hline
\rowcolor{lightgray}\cellcolor{white}5 & \textbf{1H 1659-487}\tablefootmark{22} & \xebf{2.56}{-08} $\dagger$ & $\mathbf{V=14.7\pm0.1}$ & $\mathbf{r\sim19.9\pm0.1}$ & $\mathbf{1.2}$ &$\mathbf{42.14\pm0.01}$ &\textbf{\cite{Buxton2012,Shahbaz2001,Hynes2004,Hynes2003a}} 
\\
\rowcolor{lightgray}\cellcolor{white}& \textbf{(GX 339-4)} & & & & & & 
\\
\hline
4 & 4U 1755-338 & \xe{1.17}{-09} & $V\sim18.456$ & $R>21.555$ & $0.624^\S$ &$\sim4.4$ &\cite{Giacconi1974,Mason1985} 
\\
& (V4134 Sgr) & & & & & & \cite{Wachter1998,White1984} 
\\
\rowcolor{lightgray}\cellcolor{white}3 & \textbf{4U 1543-475}\tablefootmark{23} & \xebf{9.42}{-08} & $\mathbf{V=14.856}$ & $\mathbf{I=16.1\pm0.1}$ & $\mathbf{0.5}$ &$\mathbf{26.79377\pm0.00007}$ &\textbf{\cite{Park2004,Pedersen1983a,Orosz1998}} 
\\
\rowcolor{lightgray}\cellcolor{white} & \textbf{(IL Lup)} & & & & & & \textbf{\cite{Chevalier1992,Orosz2003}}
\\
\hline
2 & 4U 1630-472\tablefootmark{24} & \xe{2.65}{-08} $\dagger$ & $K=16.1^\ast$ & $ $ & $4.2$ &$ $ &\cite{Augusteijn2001,Callanan2000} \\
& (Nor X-1) & & & & & & \\
\hline
1 & Cen X-2 & \xe{1.58}{-07} & & $ $ & $1.278^\S$ &$ $ & \cite{Brocksopp2001}\\
\hline

\end{longtable}
\begin{flushleft}
\tablefoot{\\
\textbf{Dynamical BHs are highlighted in grey boxes.}\\
 $\ast$~These values were not transformed in the AB system.\\
 $\S$~This values are estimated with the \cite{Schlafly2011} map of dust emission. Values for targets within $\vert b \vert<5$\deg~are rough estimates.\\
$\dagger$~These peak X-ray fluxes were provided by the ASM/RXTE teams at MIT and at the RXTE SOF and GOF at NASA's GSFC.\\
$\ddagger$~This peak X-ray flux was obtained from the MAXI SSC light curve.\\
\tablefoottext{1}{For SWIFT J174510.8-262411, we have used a power-law $\Gamma$=1.53 \citep{Tomsick2012}.}\\
\tablefoottext{2}{For SWIFT J1539.2-6227, we have used a power-law $\Gamma$=2.15 \citep{Krimm2009}.}\\
\tablefoottext{3}{For XTE J1818-245, we have used a power-law $\Gamma$=2.44 \citep{Markwardt2005a}. The quiescent magnitude was taken $\sim$72 days after the peak of the outburst, probably not in real quiescence \citep{CadolleBel2009}.}\\
\tablefoottext{4}{For SWIFT J1753.5-0127, we have used a power-law $\Gamma$=2.11 \citep{Morris2005}.}\\
\tablefoottext{5}{For IGR J17091-3624, we have used a power-law $\Gamma$=1.60 \citep{Kennea2007}.}\\
\tablefoottext{6}{For XTE J1720-318, we have used a power-law $\Gamma$=2.70 \citep{Chaty2006}.}\\
\tablefoottext{7}{For XTE J1908+094, we have used a power-law $\Gamma$=1.55 \citep{Woods2002}.}\\
\tablefoottext{8}{SAX J1711.6-3808: The infrared magnitudes are from a star within the X-ray error circle in a crowded field, but it is not confirmed as the true counterpart \cite{Wang2014}.}\\
\tablefoottext{9}{For XTE J1650-500, we have used a power-law $\Gamma$=1.66 \citep{Tomsick2004}.}\\
\tablefoottext{10}{For XTE J1118+480, we have used a power-law $\Gamma$=1.73 \citep{Brocksopp2010}. The peak X-ray flux is taken from the 2005 outburst \citep{Brocksopp2010}.}\\
\tablefoottext{11}{For XTE J1859+226, we have used a power-law $\Gamma$=1.70 \citep{Markwardt1999a}.}\\
\tablefoottext{12}{GRS 1739-278: The peak X-ray flux is taken from the MAXI SSC light curve of the 2014 outburst.}\\
\tablefoottext{13}{For GRS 1730-312, we have used a power-law $\Gamma$=3.88 \citep{Borozdin1995}.}\\
\tablefoottext{14}{GRO J1655-40 (Nova Sco. 1994): The peak X-ray flux is taken from the 2005 outburst, the brightest one so far \citep{Brocksopp2006}.}\\
\tablefoottext{15}{For GRS 1009-45 (N. Vel 1993), we have used a power-law $\Gamma$=1.57 \citep{Borozdin1993a}.}\\
\tablefoottext{16}{GRS J1915+105: The peak X-ray flux was obtained from the maximum value in the XTE/ASM light curves (MJD=50317.68034).}\\
\tablefoottext{17}{For GRS 1124-684 (N. Mus 1991), we have used a power-law $\Gamma$=2.50 \citep{Sunyaev1992}.}\\
\tablefoottext{18}{For EXO 1846-031, we have used a power-law $\Gamma$=1.43 \citep{Parmar1993}.}\\
\tablefoottext{19}{For SLX 1746-331, we have used a power-law $\Gamma$=1.70 \citep{Homan2003}. The peak X-ray flux is taken from the 2003 outburst.}\\
\tablefoottext{20}{For H 1743-322, we have used a power-law $\Gamma$=1.49 \citep{Markwardt2003}. The X-ray data is taken from the 2003 outburst \citep{Markwardt2003}.}\\
\tablefoottext{21}{3A 1524-617 (KY~TrA): The orbital period estimate is very uncertain.}\\
\tablefoottext{22}{1H J1659-487 (GX~339-4): The peak X-ray flux is taken from the 2002 outburst \citep{Homan2005}. The peak outburst magnitude was obtained from \cite{Buxton2012} (MJD=54132).}\\
\tablefoottext{23}{4U 1543-475 (IL Lup): The peak X-ray flux is taken from the 2002 outburst \citep{Park2004}.}\\
\tablefoottext{24}{4U 1630-472 (Nor X-1): The peak X-ray flux is taken from the 2003 outburst \citep{Tomsick2005}, the brightest one detected so far.}\\
}
\end{flushleft}
}
\end{landscape}
\renewcommand{\thefootnote}{\arabic{footnote}}
\renewcommand{\arraystretch}{1.}

 \renewcommand{\arraystretch}{1.15}
\begin{landscape}
\scriptsize{
\begin{center}
\centering
\begin{longtable}{l|llll|lll|l}
\caption[A census of BHTs (III). Photometry in quiescence]{\textbf{The dynamical BHs: Photometric parameters in quiescence}\\
Columns: (1) Name of the dynamical black hole used hereafter, (2--8) quiescent magnitudes 
reported in different bands, (9) references.\\
}
\label{tab:census3}
\\
\hline
 & \multicolumn{4}{c|}{Optical bands} & \multicolumn{3}{c|}{NIR bands} & \\ 
NAME & Blue    & Green   & Red     & Infrared     &  $J$  & $H$ & $K/\ks$ & Ref.\\     
 & (AB mag) & (AB mag) & (AB mag) & (AB mag) & (AB mag) & (AB mag) & (AB mag) &  
\\
\hline
\hline
\multirow{2}{*}{\textit{Swift}\,J1357.2--0933$^\dagger$} & $u'=23.12\pm0.04$ & $g'=22.3\pm0.4$ & $r'=21.5\pm0.4$ & $i'=21.2\pm0.4$ 
   & $J=19.17\pm0.05$ & $H=18.68\pm0.05$ & $\ks=18.14\pm0.05$
   & \cite{Shahbaz2013} 
\\ 
                                         & & $g'=22.29\pm0.08$ & $r'=21.71\pm0.08$ & & & &
   & \cite{MataSanchez2015,ArmasPadilla2014a}
\\
\hline
XTE\,J1650--500 & & $V\sim24$ & $R\sim22$ & $I>19.3$ & & &
   &  \cite{Garcia2002,Curran2012}
\\ 
\hline
\multirow{2}{*}{XTE\,J1118+480} & $ B=20.04 $ & $ V=19.61 $ & $R=19.05$ & & $J=19.52$ & $H=19.1$ & $\ks=18.45$
   & \cite{Gelino2006}
\\
               &      & $g'=21.0\pm0.3$ &  & $i'=19.0\pm0.2$ & & &
   & \cite{Shahbaz2005}
\\ 
\hline
\hline
XTE\,J1859+226 &  & $V=23.25\pm0.09$ & $R=22.54\pm0.07$ & & & &
   & \cite{Zurita2002}
\\ 
(V406\,Vul) & & & & & $J\sim20.71$ & $H\sim20.59$ & $\ks\sim20.65$ & \cite{Gelino2010}
\\
\hline
SAX\,J1819.3-2525 & $B\sim13.91\pm0.04$ & $V\sim13.66\pm0.03$ & & $I\sim13.52\pm0.03$ & $J=14.86\pm0.05$ & $H\sim14.23\pm0.05$ & $K\sim12.83\pm0.05^\ast$
   & \cite{MacDonald2014}
\\
(V4641\,Sgr) & & $V=13.96\pm0.02$ & $R=13.73\pm0.02$ & &  &  & 
   & \cite{Orosz2001} 
\\
\hline
\multirow{3}{*}{XTE\,J1550--564} & & $V=22.0\pm0.2$ & $i=19.04\pm0.03$ & & $H=17.6\pm0.1$ & &
   & \cite{Calvelo2010,Jain2001}
\\ 
   & & $B>23.8\pm0.1$ & $V=22.0\pm0.4$ & & & & 
   & \cite{Orosz2002}
\\
   & &       & $R\sim22$ & & $J=19.24\pm0.04$ & & $\ks=18.00\pm0.04$ 
   & \cite{Orosz2011a} 
\\
\hline
GRO\,J1655--40 & $B\sim18.5$ & $V\sim17.5$ & & $I\sim15.4$ 
   & $J\sim14.76$ & & $\ks\sim15.12$ 
   & \cite{Greene2001}
\\
(N.~Sco\,94)   & $B\sim18.5$ & $V\sim17.3$ & $R\sim16.2$ & $I\sim15.8$ & & &
   & \cite{Orosz1997}
\\
   &           & $V=17.3\pm0.1$ & $R=16.3\pm0.1$ & $i=14.9\pm0.1$ & & &
   & \cite{VanDerHooft1998} 
\\
\hline
    
GRS\,1009-45 & & $V=21.7\pm0.2$ & $r=20.4\pm0.1$ &  &  & &
   & \cite{Hynes2003,Shahbaz1996}
\\
(N.~Vel\,93) & & & $R=21.3\pm0.2$ &&&& &\cite{Filippenko1999}
\\
\hline
\hline
GRS\,1915+105 & & & & & $J=17.52\pm0.08$\fnm[1] & $H=13.09$\fnm[1] & $K=11.4^\ast$\fnm[1] 
   & \cite{Castro-Tirado1993a,Shahbaz2008}
\\
(V1487\,Aql) &&&&& $J=18.72\pm0.2$\fnm[1] & $H=16.55\pm0.04$\fnm[1] & $\ks=15.28\pm0.03$\fnm[1]
   & \cite{Eikenberry1995}
\\
   &&&& $I\sim23.7\pm0.4$\fnm[1] &&&
   & \cite{Boeer1996}
\\
\hline
GRO\,J0422+32 &  & $V=22.3$ & $R=21.0$ & $I=20.2$ & & &
   & \cite{Garcia1996}
\\ 
(V518\,Per)   & $B=23.1\pm0.2$ & $V=22.0\pm0.1$ & $R=20.1\pm0.1$ & $I=19.8\pm0.1$ & $J=19.3\pm0.1$ & $H=19.0\pm0.2$ & $\ks=19.29\pm0.09$
   & \cite{Gelino2003}
\\
\hline
GRS\,1124-684 & $B=22.6\pm0.3$ & $V=22.62\pm0.03$ & $R=19.8\pm0.3$ & $I=19.3\pm0.2$ & & $H\sim18.39$ & & \cite{King1996a,Shahbaz1997}
\\ 
(N.~Mus\,91) & $u'=21.5\pm0.4$ & $g'=20.65\pm0.05$ && $i'=19.92\pm0.04$&&&& \cite{Shahbaz2010}
\\
\hline
GS\,2023+338 & \multirow{2}{*}{$B=20.47\pm0.05$} & \multirow{2}{*}{$V=18.38\pm0.02$} & \multirow{2}{*}{$R=16.58\pm0.01$} & \multirow{2}{*}{$I=16.07\pm0.05$} & \multirow{2}{*}{$J=14.59\pm0.08$} & \multirow{2}{*}{$H=14.21\pm0.06$} & \multirow{2}{*}{$\ks=14.39\pm0.05$} & \multirow{2}{*}{\cite{Casares1993,Zurita2004}}
\\ 
(V404\,Cyg) & & & & & & & &
\\ 
\hline
GS\,2000+251 & & & $R=21.3\pm0.2$ & $I=20.2\pm0.2$ & $J=19.0\pm0.2$ & $H=18.37\pm0.03$ & $\ks=18.9\pm0.2$
   & \cite{Callanan1991,Beekman1996}
\\
(QZ\,Vul)    & & & & & & & & 
\\
\hline
\hline
GS\,1354-64 & & & $R=20.71\pm0.03$ & & & &
   & \cite{Casares2009}
\\ 
(BW\,Cir) & & $V\sim21.5$ & $R\sim20.5$ & $i\sim20.2$ & & & & \cite{Russell2015a}
\\ 
\hline
H\,1705-250 & \multicolumn{2}{c}{\multirow{2}{*}{$B+V=21.5\pm0.1^\ast$}} & \multirow{2}{*}{$R=20.9\pm0.2$} & & & &
   & \cite{Martin1995,Remillard1996}
\\
(N.~Oph\,77) &&&& $i'\sim20.5$ &&&& \cite{Yang2012}
\\
\hline
3A0620--003 & $B=19.32\pm0.04$ & $V=18.27\pm0.04$ & $R=17.12\pm0.04$ & $I=16.93\pm0.01$ & $J=16.31\pm0.04$ & $H=16.19\pm0.04$ & 
   & \cite{Cantrell2010}
\\ 
          & & & & $I=17.01\pm0.04$ & $J=16.5\pm0.1$ & $H=16.23\pm0.02$ & & \cite{Gelino2001,Froning2001}
\\
\hline
1H\,J1659-487 & & & $r=19.9\pm0.1$ & & & &
   & \cite{Shahbaz2001}
\\ 
(GX\,339--4) & & & $R=21.6\pm0.3$ & & & & & \cite{Lewis2012}
\\ 
\hline
4U\,1543--475 &  & $V=16.6\pm0.1$ & & $I=16.1\pm0.1$ & $J=16.04\pm0.05$ & &
   & \cite{Orosz1998,Buxton2004}
\\
(IL\,Lup) & $B=17.3$ & $V=16.7$ & & & & & 
   & \cite{Chevalier1992} 
\\
\hline
\hline
\end{longtable}
\begin{flushleft}
\tablefoot{
\\
$^\dagger$~Strictly speaking, \textit{Swift}\,J1357.2--0933 is not a dynamical BHTs because it lacks of detection of the secondary star but it has a robust mass function determination. 
\\
$^\ast$~These magnitudes were not transformed into the AB system.\\
\tablefoottext{1}~Since this object has not reached the quiescent level again, the IR magnitudes are values in outburst. The $I$-band magnitude reported by \cite{Boeer1996} is estimated from a tentative optical counterpart in the field.
}
\end{flushleft}
\end{center}
}
\end{landscape}
\renewcommand{\thefootnote}{\arabic{footnote}}
\renewcommand{\arraystretch}{1.}

 \renewcommand{\arraystretch}{1.15}
\begin{landscape}
\scriptsize{
\begin{longtable}{l l lllllll l}
\caption[A census of BHTs (IV): Dynamical parameters]{\textbf{The dynamical BHs: Binary parameters}
Columns: (1) Name of the dynamical black hole used hereafter, (2) spectral type of the companion star, (3) orbital period, (4) secondary star's radial velocity amplitude, (5) mass function, (6) mass of the black hole, (7) binary mass ratio, (8) inclination of the system and (9) references.
} 
\label{tab:census4}
\\
\hline
(1) & (2) & (3) & (4) & (5) & (6) & (7) & (8) & (9) & (10)\\
NAME & Spectral & $P_{\rm orb}$ & $K_2$  & $f(M_1)$ & $M_1$   &$q$& $i$    & $v_{rot}\sin i$ 
& Ref.
\\
     & type     & ~(h)          & (km\,s$^{-1}$) & (\msun)  & (\msun) &   & (\deg) & (km\,s$^{-1}$)
& 
\\
\hline
\hline
\textit{Swift}\,J1357.2--0933$^\dagger$ & M2$-$4V & $2.8\pm0.3$ & $967\pm49$ & $11.0\pm2.1$ & $>8.9$ & $\sim0.04$ & $\sim90$ &
   & \cite{Corral-Santana2013,MataSanchez2015}$^\dagger$
\\ 
\hline
XTE\,J1650--500 & $\sim$K4V & $7.69\pm0.02$ & $435\pm30$ & $2.7\pm0.6$ & $\leq7.3$ &  & $>47$ &
   & \cite{Orosz2004}
\\ 
\hline
\multirow{2}{*}{XTE\,J1118+480} & \multirow{2}{*}{K7--M1V} & \multirow{2}{*}{$4.078414\pm0.000005$} 
   & \multirow{2}{*}{$709\pm1$} & \multirow{2}{*}{$6.27\pm0.04$} 
   & \multirow{2}{*}{$6.9-8.2$} & \multirow{2}{*}{$0.024\pm0.009$} & \multirow{2}{*}{$68-79$} & \multirow{2}{*}{$96^{+3}_{-11}$}
   & \cite{Khargharia2013,Torres2004}
\\
   & & & & & & & &
   & \cite{GonzalezHernandez2008,Calvelo2009}
\\ 
\hline
\hline
XTE\,J1859+226 & \multirow{2}{*}{$\sim$K5V} & \multirow{2}{*}{$6.58\pm0.05$} & \multirow{2}{*}{$541\pm70$} & \multirow{2}{*}{$4.5\pm0.6$} & \multirow{2}{*}{$>5.42$} & & \multirow{2}{*}{$<70$} &
   & \multirow{2}{*}{\cite{Corral-Santana2011}}
\\ 
(V406\,Vul) & & & & & & & & &
\\
\hline
SAXJ1819.3-2525 & \multirow{2}{*}{B9III} & \multirow{2}{*}{$67.6152\pm0.0002$} & \multirow{2}{*}{$211\pm3$} & \multirow{2}{*}{$2.7\pm0.1$} & \multirow{2}{*}{$6.4\pm0.6$} & \multirow{2}{*}{$0.63-0.70$} & \multirow{2}{*}{$72\pm4$} & \multirow{2}{*}{$100.9\pm0.8$}
   & \multirow{2}{*}{\cite{Orosz2001,MacDonald2014}}
\\
(V4641\,Sgr) & & & & & & & & &
\\
\hline
XTE\,J1550--564 & K2--K4IV & $37.0088\pm0.0001$ & $363\pm6$ & $7.65\pm0.38$ & $7.81-15.6$ & $\approx0.03$ & $75\pm4$ & $55\pm5$
   & \cite{Orosz2002,Orosz2011a}
\\ 
\hline
GRO\,J1655--40 & \multirow{2}{*}{F6IV} & \multirow{2}{*}{$62.920\pm0.003$} & \multirow{2}{*}{$226.1\pm0.8$} & \multirow{2}{*}{$2.73\pm0.09$} & \multirow{2}{*}{$6.0\pm0.4$} & \multirow{2}{*}{$0.42\pm0.03$} & \multirow{2}{*}{$69\pm2$} & \multirow{2}{*}{$86^{+3.3}_{-3.6}$}
   & \cite{Orosz1997,VanDerHooft1998,Shahbaz1999,Shahbaz2003}
\\
(N.~Sco\,94) & & & & & & & & 
   & \cite{Beer2002,GonzalezHernandez2008a}
\\
\hline
GRS\,1009-45 & \multirow{2}{*}{K7--M0V} & \multirow{2}{*}{$6.8449\pm0.0003$} & \multirow{2}{*}{$475\pm6$} & \multirow{2}{*}{$3.2\pm0.1$} & \multirow{2}{*}{$\gtrsim4.4$} & \multirow{2}{*}{$0.14\pm0.02$} & \multirow{2}{*}{$37-80$} &  & \multirow{2}{*}{\cite{Filippenko1999,Shahbaz1996}}
\\
(N.~Vel\,93) & & & & & & & & &
\\
\hline
\hline
GRS\,1915+105 & \multirow{2}{*}{K1--5III} & \multirow{2}{*}{$812\pm4$} & \multirow{2}{*}{$126\pm1$} & \multirow{2}{*}{$7.0\pm0.2$} & \multirow{2}{*}{$10.1\pm0.6$} & \multirow{2}{*}{$0.042\pm0.024$} & \multirow{2}{*}{$66\pm2$} & \multirow{2}{*}{$21\pm4$}
    & \multirow{2}{*}{\cite{Harlaftis2004,Steeghs2013,Fender1999}}
\\
(V406\,Vul) & & & & & & & & &
\\
\hline
\multirow{2}{*}{GRO\,J0422+32} & \multirow{2}{*}{M4--5V} 
    & \multirow{2}{*}{$5.094185\pm0.000005$} & \multirow{2}{*}{$378\pm16$} 
    & \multirow{2}{*}{$1.19\pm0.02$} & \multirow{2}{*}{$2-15$} & 
    \multirow{2}{*}{$0.11_{-0.02}^{+0.05}$} & \multirow{2}{*}{$10-50$} & \multirow{2}{*}{$90_{-27}^{+22}$}   
    & \cite{Webb2000,Casares1995c}
\\
 && &&&&&&
    & \cite{Beekman1997,Harlaftis1999}
\\ 
\hline
GRS\,1124-684 & \multirow{2}{*}{K3--5V} & \multirow{2}{*}{$10.38254\pm0.00007$} 
   & \multirow{2}{*}{$406\pm7$} & \multirow{2}{*}{$3.0\pm0.2$} & \multirow{2}{*}{$3.8-7.5$} 
   & \multirow{2}{*}{$0.09-0.17$} & \multirow{2}{*}{$39-65$} & \multirow{2}{*}{$106\pm13$}
   & \cite{Orosz1996,Shahbaz1997,Gelino2001a}
\\ 
(N.~Mus\,91) &&&&&&& &
 &\cite{Gelino2001,Casares1997}
\\
\hline
GS\,2023+338 & \multirow{2}{*}{K3III} & \multirow{2}{*}{$155.311\pm0.002$} 
   & \multirow{2}{*}{$208.5\pm0.7$} & \multirow{2}{*}{$6.08\pm0.06$} 
   & \multirow{2}{*}{$9^{+0.2}_{-0.6}$} & \multirow{2}{*}{$0.055-0.064$} & \multirow{2}{*}{$52-80$} & \multirow{2}{*}{$39\pm1$}
   & \cite{Wagner1992,Casares1992,Casares1993}
\\ 
(V404\,Cyg) & & & & & & & & & \cite{Casares1994,Shahbaz1994b,Khargharia2010}
\\
\hline
  GS\,2000+251 & \multirow{2}{*}{K3--7V} & \multirow{2}{*}{$8.258095\pm0.000005$} 
  & \multirow{2}{*}{$520\pm5$} & \multirow{2}{*}{$5.0\pm0.1$} 
  & \multirow{2}{*}{$5.5-8.8$} & \multirow{2}{*}{$0.04\pm0.01$} & \multirow{2}{*}{$54-60$} & \multirow{2}{*}{$86\pm8$}
  & \cite{Charles1991,Casares1995a}
\\ 
 (QZ\,Vul) && &&&&&& & \cite{Harlaftis1996,Ioannou2004}
\\
\hline
\hline
 GS\,1354-64 & \multirow{2}{*}{G5III} & \multirow{2}{*}{$61.068\pm0.002$} 
   & \multirow{2}{*}{$279\pm5$} & \multirow{2}{*}{$5.7\pm0.3$} & \multirow{2}{*}{$\geq7.6\pm0.7$} 
   & \multirow{2}{*}{$0.12^{+0.03}_{-0.04}$} & \multirow{2}{*}{$\leq79$} & \multirow{2}{*}{$69\pm8$}
   & \multirow{2}{*}{\cite{Casares2004,Casares2009}}
\\ 
(BW\,Cir) & & & & & & & & &
\\
\hline
H\,1705-250 & \multirow{2}{*}{K3--M0V} & \multirow{2}{*}{$12.51\pm0.03$} 
   & \multirow{2}{*}{$448\pm4$} & \multirow{2}{*}{$4.9\pm0.1$} & \multirow{2}{*}{$4.9-7.9$} 
   & \multirow{2}{*}{$\leq0.053$} & \multirow{2}{*}{$48-80$} & \multirow{2}{*}{$\leq79$}
   & \cite{Remillard1996,Filippenko1997}
\\
(N.~Oph\,77) &&&&&&&&& \cite{Harlaftis1997,Martin1995}
\\
\hline
\multirow{2}{*}{3A0620--003} & \multirow{2}{*}{K2--7V} & \multirow{2}{*}{$7.7523372\pm0.0000002$} 
  & \multirow{2}{*}{$437\pm2$} & \multirow{2}{*}{$2.79\pm0.04$} & \multirow{2}{*}{$6.6\pm0.3$} 
  & \multirow{2}{*}{$0.074\pm0.006$} & \multirow{2}{*}{$51.0\pm0.9$} &
  & \cite{Cantrell2010,Orosz1994}
\\
 &&&&&&&&& \cite{GonzalezHernandez2010}
\\ 
\hline
1H\,J1659-487 & \multirow{2}{*}{$>$GIV} & \multirow{2}{*}{$42.14\pm0.01$} & \multirow{2}{*}{$>317\pm10$} & \multirow{2}{*}{$5.8\pm0.5$} & \multirow{2}{*}{$>6$} & \multirow{2}{*}{$\leq0.125$} & &
   & \multirow{2}{*}{\cite{Hynes2003a,Hynes2004,Munoz-Darias2008}} 
\\
(GX\,339-4) & & & & & & & & &
\\
\hline
4U\,1543--475 & \multirow{2}{*}{A2V} & \multirow{2}{*}{$26.79377\pm0.00007$} & \multirow{2}{*}{$124\pm4$} & \multirow{2}{*}{$0.25\pm0.01$} & \multirow{2}{*}{$8.4-10.4$} & \multirow{2}{*}{$3.2-4.0$} & \multirow{2}{*}{$20.7\pm1.5$} & \multirow{2}{*}{$46\pm2$} & \multirow{2}{*}{\cite{Orosz1998,Orosz2003}}
\\
(IL\,Lup) & & & & & & & & &
\\
\hline
\hline
\end{longtable}

\tablefoot{
\begin{flushleft}
$^\ast$~The privileged values of $M_1$ were assumed using \cite{Casares2014a}.\\
$^\dagger$~Strictly speaking, \textit{Swift}\,J1357.2--0933 is not a dynamical BHTs because it lacks of detection of the secondary star but it has a robust mass function determination.\\
\end{flushleft}
}
}
\end{landscape}
\renewcommand{\thefootnote}{\arabic{footnote}}
\renewcommand{\arraystretch}{1.}

\end{longtab}

\begin{acknowledgements}
      We thank the anonymous referee for useful comments. We acknowledge financial support from CONICYT-Chile grants FONDECYT Postdoctoral Fellowship 3140310 (JMC-S), FONDECYT 1141218 (FEB), Basal-CATA PFB-06/2007 (JMS-C, FEB), "EMBIGGEN" Anillo ACT1101 (FEB), the Ministry of Economy, Development, and Tourism's Millennium Science Initiative through grant IC120009, awarded to The Millennium Institute of Astrophysics, MAS (FEB) and the Spanish Ministerio de Econom\'ia y Competitividad (MINECO) under grant AYA\,2013-42627 (JC, TMD, IGMP). TMD acknowledges hospitality during his 2015 visit to IA-PUC. The finding charts were obtained from: the ESO Science Archive Facility under request number jcorral-160882; the GTC Public Archive at CAB (INTA-CSIC) and the Isaac Newton Group and LCOGT archives which are maintained as part of the CASU Astronomical Data Centre at the Institute of Astronomy, Cambridge. We are thankful to Danny Steeghs and Manuel A.~P.~Torres for gently provide us some of the finding charts and Jorge Andr\'es Perez Prieto for his help with the creation of the web. 
      We have used the web applications of the \href{http://heasarc.gsfc.nasa.gov/ftools/}{FTOOLS} \citep{Blackburn1995} and \href{http://heasarc.gsfc.nasa.gov/cgi-bin/Tools/w3pimms/w3pimms.pl}{PIMMS} \citep{Mukai1993} software to make the transformation of the X-ray fluxes. Some peak X-ray fluxes were provided by the ASM/RXTE teams at MIT and at the RXTE SOF and GOF at NASA's GSFC. This research has made use of the MAXI data provided by RIKEN, JAXA and the MAXI team \citep{Matsuoka2009}.
\end{acknowledgements}

\bibliographystyle{aa} 
\bibliography{catalogue.bib} 

\end{document}